\documentclass{aa}

\usepackage{graphicx}
\usepackage[utf8]{inputenc}
\usepackage{amssymb}
\usepackage{amsmath}
\usepackage{graphicx} 
\usepackage{enumitem}
\usepackage{url}
\usepackage{multirow}
\usepackage{comment}
\setlist{nosep} %
\usepackage[colorlinks=true,linkcolor=blue,citecolor=blue]{hyperref}
\usepackage{ulem}

\usepackage[varg]{txfonts}
\newcommand{\simpropto}{\mathrel{\vcenter{
  \offinterlineskip\halign{\hfil$##$\cr
    \propto\cr\noalign{\kern2pt}\sim\cr\noalign{\kern-2pt}}}}}

\newcommand{\GRB}{GRB~221009A}

\newcommand{\revone}[1]{#1}
\newcommand{\revtwo}[1]{#1}

\defcitealias{Blandford1976}{BM76}
\defcitealias{Beloborodov2002}{Be02}
\defcitealias{Ravasio2024}{R24}

\begin{document}

   \title{A self-consistent explanation of the MeV line in GRB~221009A unveils a dense circum-stellar medium}
   \titlerunning{MeV line in GRB~221009A unveils dense CSM}

   \author{{Om Sharan} Salafia\thanks{om.salafia@inaf.it}\inst{1,2}, 
           {Annalisa} Celotti\inst{3,1,4,5}, 
           {Emanuele} Sobacchi\inst{6,7}, 
           {Lara} Nava\inst{1}, 
           {Gor} Oganesyan\inst{6,7}, 
           {Giancarlo} Ghirlanda\inst{1}, 
           {Stella} Boula \inst{1}, 
           {Maria Edvige} Ravasio\inst{8,9,1}, 
           {Gabriele} Ghisellini\inst{1} 
   }

   \institute{INAF -- Osservatorio Astronomico di Brera, via Emilio Bianchi 46, I-23807 Merate (LC), Italy
        \and
             INFN -- sezione di Milano-Bicocca, Piazza della Scienza 3, I-20126 Milano (MI), Italy
        \and
             SISSA, via Bonomea 265, I-34136 Trieste, Italy
        \and 
             INFN -- sezione di Trieste, via Valerio 2, I-34127, Trieste, Italy
        \and
            IFPU -- Institute for Fundamental Physics of the Universe, via Beirut 2, I-34151 Trieste, Italy
        \and 
             Gran Sasso Science Institute, viale F. Crispi 7, L'Aquila, I-67100, Italy
        \and 
            INFN -- Laboratori Nazionali del Gran Sasso, L'Aquila, I-67100, Italy
        \and
             Institute of Space Sciences (ICE, CSIC), Campus UAB, Carrer de Can Magrans s/n, E-08193, Barcelona, Spain
        \and 
            Department of Astrophysics/IMAPP, Radboud University, PO Box 9010, 6500 GL Nijmegen, The Netherlands
             }

    \authorrunning{O.~S.\ Salafia et al.}
  
   \date{Received 18/12/2025; accepted 16/06/2026}


 \abstract{GRB~221009A has been the brightest gamma-ray burst (GRB) observed to date, and its afterglow has been characterized with unprecedented detail at TeV energies by LHAASO. Quite puzzlingly, it is also the most energetic GRB known. Among the riddles posed by this mysterious source, however, the sheer energetics are hardly the most intriguing: an unprecedented, narrow, luminous emission line at around 10 MeV has been uncovered by a detailed spectral analysis of \textit{Fermi}/GBM data immediately following the brightest peak in the GRB prompt emission and the peak of the TeV afterglow. As noted in the discovery article, the temporal evolution of the line properties can be explained as being due to high-latitude emission from a geometrically thin, relativistically expanding shell where annihilation of a large number of electron-positron pairs took place. We show that this interpretation yields stringent constraints on the properties of such shell, that point to a process that happens at radii typical of external shocks. We then demonstrate that the shell could have been the blastwave associated with the GRB precursor, with the line arising after pair loading of such blastwave as it was illuminated by the bright and hard radiation of the GRB main event. The scenario, which also explains the abrupt initial rise of the LHAASO afterglow, requires the progenitor of the GRB to have been surrounded by a circum-stellar medium (CSM) extending out to a few $10^{15}\,\mathrm{cm}$, with a density $n_\mathrm{ext}\sim 10^{8}-10^{9}\,\mathrm{cm^{-3}}$ reminiscent of those found from studies of Type IIn supernovae. \revtwo{The consequences of such a CSM on the dynamics and emission of the external shock are yet to be fully explored. If future, more detailed work will confirm the compatibility of the GRB 221009A afterglow emission with our scenario, this will} provide a precious clue to the nature of the progenitor of this peculiar GRB, which could also be present in other bursts that feature a long quiescence followed by a bright emission episode with a hard spectrum.}


   \maketitle

   %

\section{Introduction}

Despite decades of research and a growing sample of well-characterized examples, gamma-ray bursts (GRBs, \citealt{Kumar2015}) remain among the most elusive astronomical sources. The association of some members of this class with core-collapse supernovae \citep{Galama1998} and studies of their host galaxies \citep{Vergani2015} clearly point to massive star progenitors, but the actual properties of the progenitor stars, and how these are linked to the observed GRB properties (such as energetics, duration, spectral features) remain poorly understood. In part, this is a consequence of the fact that the mechanism behind GRB `prompt' emission (i.e.\ the initial, highly variable emission that is detected through space-based hard-X-ray and soft gamma-ray observatories) is still not well constrained. Even the longer-lasting `afterglow', whose evolution is less erratic, and whose underlying mechanisms are better understood \citep{Panaitescu2000}, often presents features that are not easily interpreted within the standard picture.

A particularly puzzling example is the extremely bright GRB~221009A (the brightest ever recorded to date -- \citealt{Malesani2025}), which was detected on October 9, 2022 by virtually all X-ray and gamma-ray detectors on spacecraft, including the Gamma-ray Burst Monitor (GBM) and the Large Area Telescope (LAT) onboard \textit{Fermi} \citep{Lesage2023,Axelsson2024}, the Konus detector onboard \textit{WIND} \citep{Frederiks2023} and the GECAM-C cubesat \citep{Zhang2024_observation}. Throughout this article, we take the \textit{Fermi}/GBM trigger time, 13:16:59.9 universal time (UT), as our observer-frame reference time. The gamma-ray light curve featured an initial pulse of about 10 s, followed by a long apparent quiescence of about 180 s, after which a highly variable and bright emission event (showing multiple peaks, some of which saturated the GBM detectors) took place, lasting hundreds of seconds. 

At a UT time corresponding to 230 seconds after the \textit{Fermi}/GBM trigger, the Water Cherenkov Detector Array (WCDA) of the Large High Altitude Air Shower Observatory (LHAASO, \citealt{LHAASO2023}) detected a very high-significance, transient emission at TeV photon energies, coming from a direction consistent with the location of the GRB. The TeV light curve, characterized with unprecedented detail, showed a very sharp rise of a few seconds, followed by a slightly gentler rise of about 10 seconds, and then a smooth decay (see bottom panel of Fig. \ref{fig:lc_and_line_props}).

At around 280 s, shortly after the brightest peak of the prompt emission and the onset of the TeV afterglow, \citet{Ravasio2024} (hereafter \citetalias{Ravasio2024}) discovered an unprecedented, highly significant, narrow emission feature with a luminosity of about $10^{50}\,\mathrm{erg\,s^{-1}}$ and a central photon energy slightly in excess of 10 MeV. A detailed spectral analysis showed an evolution of the line properties towards lower luminosity and photon energy during the subsequent minute, after which the feature could not be detected anymore. A statement by the \textit{Fermi}/GBM team \citep{Burns2024} confirmed the presence of the feature in the data and excluded an instrumental artifact. The feature was further analysed and characterized in subsequent works \citep{Axelsson2024,Zhang2024_observation}, and a hint of a related feature was reported in Konus-\textit{WIND} data \citep{Frederiks2023}.

In this paper, we propose a scenario to explain the emergence of the line, its timing, its properties and their evolution. As we will demonstrate, the scenario provides a self-consistent explanation of various features of the prompt emission and afterglow of this GRB, and yields a clue about the nature of its progenitor.
In Section \ref{sec:HLE_interpretation} we show that the evolution of the line properties can be interpreted as high-latitude emission (HLE) from a shell in relativistic expansion within which mass-annihilation of a large number of electron-positron ($e^-e^+$) pairs took place, as already proposed by \citetalias{Ravasio2024}, \citet{Zhang2024_observation,Zhang2024_implications}, and \citet{Peer2024} (see \citealt{Liu2025} for an alternative scenario). We show that this interpretation poses quite stringent constraints on the number of annihilating pairs and on the bulk Lorentz factor of the shell, and also on the observer-frame time at which the annihilation started. In Section \ref{sec:Scenario} we describe our proposed scenario for the production of the required number of pairs and their mass-annihilation and demonstrate its general feasibility. In Section \ref{sec:pair_enrichment} we describe our detailed model of the process that leads to the pair enrichment and acceleration of the pair-loaded shell, and derive the constraints on the model parameter space imposed by the HLE interpretation. Finally, in Section \ref{sec:discussion} we discuss our results and their consequences on the properties of the progenitor. We provide a compact summary and our conclusions in Section \ref{sec:conclusions}.

Throughout this article, we employ the usual notation $Q_x = Q_\mathrm{cgs}/10^x$, where $Q$ is any quantity and $Q_\mathrm{cgs}$ is its value in cgs units. Primed quantities are measured in the fluid frame. When a cosmology is needed, we adopt a flat $\Lambda$CDM model with \citet{Planck2020} parameters. The redshift of GRB~221009A, determined spectroscopically, is $z=0.151$ \citep{Malesani2025}.

\section{Interpreting the line evolution as due to high latitude emission}\label{sec:HLE_interpretation}

\begin{figure*}
 \centering
 \includegraphics[width=0.9\textwidth]{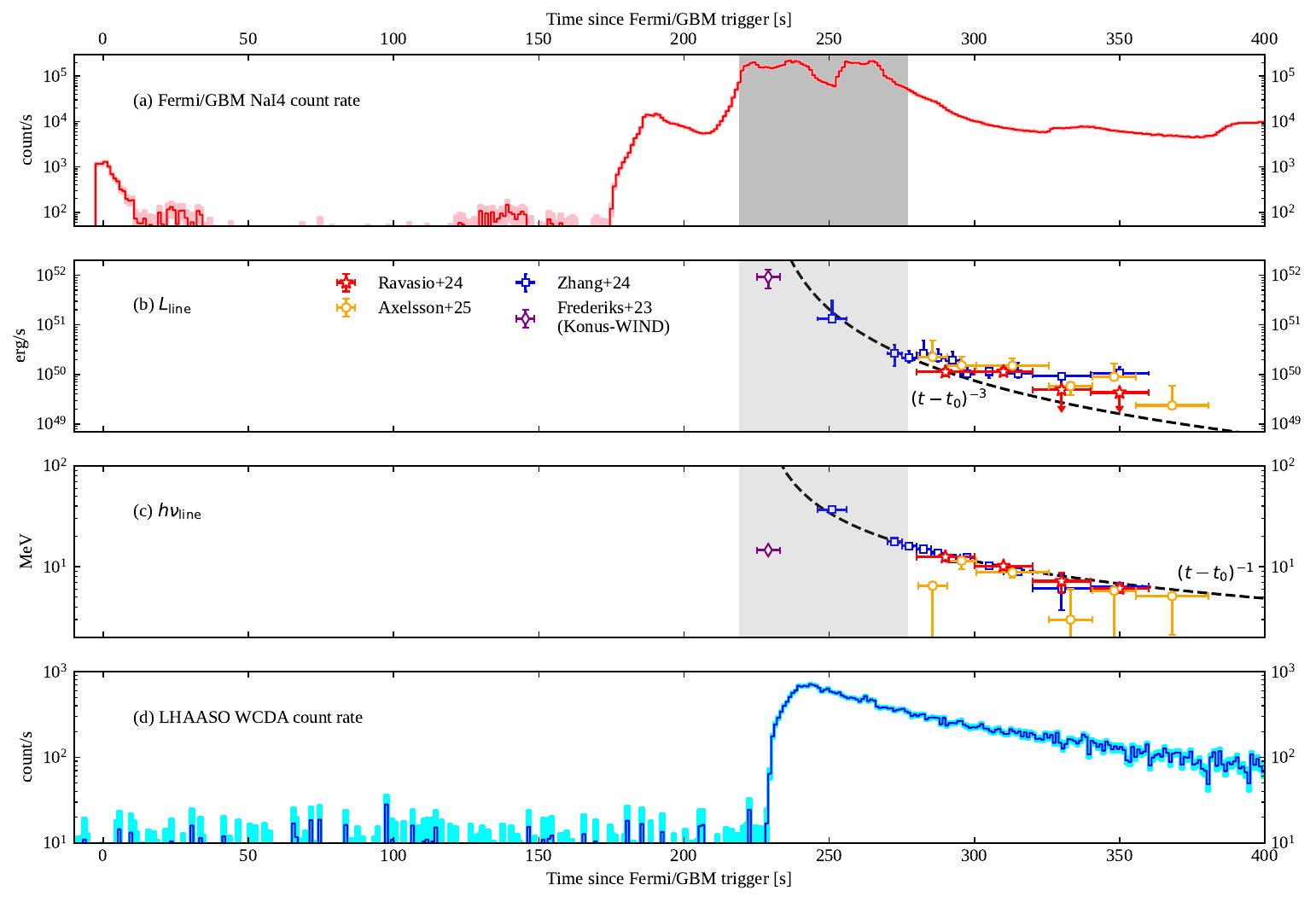}
 \caption{GRB~221009A light curves and narrow line properties. \textbf{Panel (a):} background-subtracted count rate light curve of the fourth sodium-iodide (NaI) detector of \textit{Fermi}/GBM (red histogram, sensitive to emission in the 8 keV -- 1 MeV band). The pink shading shows the one-sigma-equivalent uncertainty, assuming Poisson-distributed background counts. The gray-shaded area shows the `Bad Time Interval' (BTI) during which the GBM detectors were affected by saturation. \textbf{Panel (b):} Gaussian line luminosity evolution. Error bars show the estimated Gaussian line luminosity according to the analysis of GBM data reported in \citetalias{Ravasio2024} (red stars), \citeauthor{Axelsson2024} (\citeyear{Axelsson2024}, orange circles), \citet[][blue squares]{Zhang2024_observation}, and that of the tentatively detected line in Konus-\textit{WIND} data \citep[][purple diamond]{Frederiks2023}. The BTI time range (only relevant to GBM) is reported with a lighter gray shading. \textbf{Panel (c):} same as panel (b), but for the line central photon energy. \textbf{Panel (d):} LHAASO WCDA count rate light curve \citep{LHAASO2023}. The cyan shading shows the one-sigma-equivalent uncertainty, assuming Poisson-distributed background counts.}
 \label{fig:lc_and_line_props}
\end{figure*}

Observationally, as shown\footnote{The orange error bars in Fig.\ \ref{fig:lc_and_line_props} reflect the results reported in table 9 of \citet{Axelsson2024}, but we note that these seem to conflict with the visualization in their figure 8, where the line evolution seems rather smooth and monotonic.} in the two middle panels of Figure \ref{fig:lc_and_line_props}, the temporal evolution of the line properties is consistent with $L_\mathrm{line}\propto (t-t_0)^{-3}$ and $h\nu_\mathrm{line}\propto (t-t_0)^{-1}$ if the reference time $t_0$ is taken close to the peak of the GRB 221009A prompt emission, $220\lesssim t_0/\mathrm{s}\lesssim 240$ \citepalias{Ravasio2024}. This is confirmed also by the independent analysis of \citet{Zhang2024_observation}. While the latter analysis raises reasonable concerns because it makes use of formally saturated \textit{Fermi}/GBM data \citep{Lesage2023}, the \textit{Fermi}/GBM Collaboration considers it acceptable, at least in part \citep{Burns2024}. In addition, \citet{Frederiks2023} reported about a possible hint at the presence of the line in Konus-\textit{WIND} spectra covering the interval from 225 to 233 s.  Differently from their assessment, we find that the photon energy and luminosity of such tentative line are not consistent with the extrapolation of the aforementioned power laws (Fig.\ \ref{fig:lc_and_line_props}, panels b and c).

This temporal evolution can be interpreted \citepalias{Ravasio2024} as being due to the HLE effect that arises when a spherical source produces a brief flash of radiation while expanding at relativistic speed. If the emission duration of the flash satisfies $\Delta t_\mathrm{em}\ll r/c$, where $r$ is the emission radius, then the temporal evolution of the observed radiation is set by the delayed arrival of photons emitted at increasing latitudes (and with decreasing Doppler factors) with respect to the source axis parallel to the line of sight \citep{Kumar2000}. In Appendix \ref{sec:HLE_derivation} we give a derivation that clarifies some aspects of this effect with respect to that in the original paper.

\subsection{Requirements set by HLE in the $e^+e^-$ pair annihilation scenario}\label{sec:HLErequirements}

If a narrow emission line with a central photon energy $h\nu_\mathrm{line}^\prime$ (as measured in the rest frame of the emitting material) is produced within a geometrically thin spherical shell of radius $r$ over a time shorter than $r/c$ ($c$ being the speed of light), while the spherical shell is expanding with a bulk Lorentz factor $\Gamma\gg 1$, then the evolution of the line luminosity and photon energy with the observer-frame time $t_\mathrm{obs}$ is set by the HLE effect, and is given by (see Appendix \ref{sec:HLE_derivation})
\begin{eqnarray}
 & L_\mathrm{line}(t_\mathrm{obs}) \sim \displaystyle\frac{2E_\mathrm{rad}/t_\mathrm{ang}}{\left[ 1+\left(t_\mathrm{obs}-t_0\right)/(1+z)t_\mathrm{ang}\right]^3},
 \label{eq:L_HLE}\\
 & h\nu_\mathrm{line}(t_\mathrm{obs}) \sim \displaystyle\frac{2\Gamma h\nu_\mathrm{line}^\prime/(1+z)}{1+\left(t_\mathrm{obs}-t_0\right)/(1+z)t_\mathrm{ang}}
 \label{eq:hnu_HLE},
\end{eqnarray}
where $E_\mathrm{rad}$ is the total radiated energy, $t_0$ is the arrival time of the first photon and $t_\mathrm{ang}=r/2\Gamma^2 c$ is the `angular' time scale. In words, the HLE effect produces a nearly constant luminosity $L_\mathrm{line}\sim E_\mathrm{rad}/t_\mathrm{ang}$ and line photon energy $h\nu_\mathrm{line}\sim \Gamma h\nu_\mathrm{line}^\prime/(1+z)$ as long as $t_\mathrm{obs}-t_0<(1+z)t_\mathrm{ang}$, followed by a power law decline $L_\mathrm{line}\propto (t_\mathrm{obs}-t_0)^{-3}$ and $h\nu_\mathrm{line}\propto(t_\mathrm{obs}-t_0)^{-1}$.

\subsubsection{Constraints on radius and bulk Lorentz factor}

If the line is attributed to electron-positron pair annihilation, and its evolution is due to the HLE effect, then some stringent constraints on the properties of the thin shell can be deduced as follows.  At $t_\mathrm{obs}-t_0\gg (1+z)t_\mathrm{ang}$ we have, from Eq.\ \ref{eq:hnu_HLE},
\begin{equation}
 \frac{t_\mathrm{obs}-t_0}{(1+z)t_\mathrm{ang}}\sim\frac{2\Gamma h\nu_\mathrm{line}^\prime}{(1+z)h\nu_\mathrm{line}}=\frac{2\Gamma m_\mathrm{e}c^2}{(1+z)h\nu_\mathrm{line}},
 \label{eq:t/tang}
\end{equation}
where we assumed $h\nu_\mathrm{line}^\prime = m_\mathrm{e}c^2$. Using the luminosity evolution, Eq.\ \ref{eq:L_HLE}, we then have
\begin{equation}
L_\mathrm{line} \sim \frac{8 N_\pm \Gamma^3 m_\mathrm{e}c^3}{r}\left[\frac{2\Gamma m_\mathrm{e}c^2}{(1+z)h\nu_\mathrm{line}}\right]^{-3} = \frac{N_\pm m_\mathrm{e}c^3}{r}\left[\frac{m_\mathrm{e}c^2}{(1+z)h\nu_\mathrm{line}}\right]^{-3},
\end{equation}
where we assumed $E_\mathrm{rad}=2\Gamma m_\mathrm{e}c^2 N_\pm$, that is, an energy $2\Gamma m_\mathrm{e}c^2$ contributed by each of the $N_\pm$ annihilating pairs\footnote{This implies cold pairs and neglects three-photon annihilation, which concerns only a small fraction of the pairs -- e.g.\ \citealt{Svensson1987}.}. This allows us to express the total number of annihilating pairs as a function of measured observables, plus the radius $r$ as the only free parameter,
\begin{equation}
 N_\pm \approx 1.8\times 10^{57} r_{16} L_\mathrm{line,50}\left(\frac{h\nu_\mathrm{line}}{12.6\,\mathrm{MeV}}\right)^{-3}\left(\frac{1+z}{1.15}\right)^{-3},
 \label{eq:Npm}
\end{equation}
where the reference values for $L_\mathrm{line}$ and $h\nu_\mathrm{line}$ are based on the analysis of the first significant time bin of \citetalias{Ravasio2024}.

From Eq.\ \ref{eq:t/tang} and the definition of $t_\mathrm{ang}$, we could also derive a constraint on the bulk Lorentz factor,
\begin{equation}
 \Gamma_\mathrm{HLE} \sim \frac{r}{c(t_\mathrm{obs}-t_0)}\frac{m_\mathrm{e}c^2}{h\nu_\mathrm{line}}\approx 225\,r_\mathrm{16}\left(\frac{t_\mathrm{line}-t_0}{60\,\mathrm{s}}\right)^{-1}\left(\frac{h\nu_\mathrm{line}}{12.6\,\mathrm{MeV}}\right)^{-1},
\label{eq:GammaHLE}
\end{equation}
where $t_\mathrm{line}$ here refers to the time at which $h\nu_\mathrm{line}$ is measured (the central time of the first significant bin analysed in \citetalias{Ravasio2024} is $t_\mathrm{line}=290\,\mathrm{s}$) and we took $t_0=230$ s as reference.

As stated above, the HLE interpretation requires the emission time to be shorter than $r/c$. In the pair annihilation scenario, this in turn requires the comoving annihilation time $t^\prime_\mathrm{ann}\sim 8 n_\pm^\prime/3\sigma_\mathrm{T}c {n_\pm^\prime}^2$ (where $n_\pm^\prime$ is the comoving pair number density, and we assume cold pairs) to be shorter than $r/\Gamma c$. We can get some insight from this requirement if we express the shell comoving thickness as $\Delta r^\prime = \xi r/\Gamma$, so that the comoving volume is $V^\prime = \xi 4\pi r^3/\Gamma$. Here $\xi$ is a dimensionless constant, with typical values $\xi\sim 0.1-0.3$ for the downstream of an external shock (\citealt{Blandford1976}, hereafter \citetalias{Blandford1976}). Then $n^\prime_\pm = N_\pm/V^\prime$ and the $t^\prime_\mathrm{ann}<r/\Gamma c$ inequality becomes, using Eq.\ \ref{eq:Npm},
\begin{equation}
\begin{split}
 & \xi < \frac{3 \sigma_\mathrm{T} L_\mathrm{line}}{32\pi m_\mathrm{e}c^3 r}\left[\frac{m_\mathrm{e}c^2}{(1+z)h\nu_\mathrm{line}}\right]^{3} \approx\\
 &\approx 0.4 r_{16}^{-1} L_\mathrm{line,50}\left(\frac{h\nu_\mathrm{line}}{12.6\,\mathrm{MeV}}\right)^{-3}\left(\frac{1+z}{1.15}\right)^{-3}.
\end{split}
\end{equation}
This result shows that, were our pairs confined within the typical volume of an external shock downstream at $r\lesssim\mathrm{few}\times 10^{16}\,\mathrm{cm}$, they would annihilate fast enough for the annihilation line emission to be in the HLE regime.

\subsubsection{Constraint on the annihilation start time $t_0$}\label{sec:constraints_from_HLE_fitting}

\begin{figure}
 \centering
 \includegraphics[width=\columnwidth]{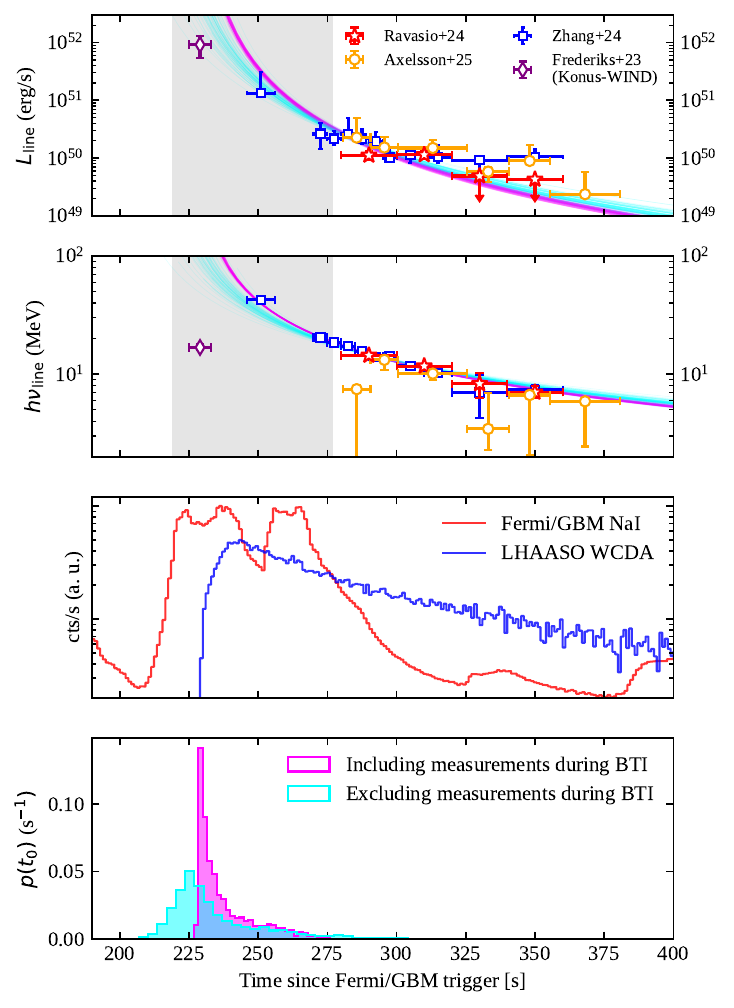}
  \caption{HLE model fit results. From top to bottom: in the first (second) panel, purple lines show the evolution of $L_\mathrm{line}$ ($h\nu_\mathrm{line}$) from one hundred posterior samples of the HLE model fitted to the evolution inferred from the observations (shown by the error bars -- see text), including data in the BTI (grey shaded area). Cyan lines show the corresponding result when such data is excluded. The third panel shows the \textit{Fermi}/GBM and LHAASO WCDA light curves as temporal references. In the last panel, the purple and cyan histograms show the posterior probability of the HLE start time $t_0$ inferred from the same two model fitting procedures.}
 \label{fig:HLE_fit_results}
\end{figure}

\begin{table}[t]
 \caption{Summary of HLE parameter constraints from model fitting.}
 \centering\small
 \begin{tabular}{lll}

 Parameter & Mean $\pm$ st. dev. & 90\% C.R.\\
 \hline\hline
  $x_0 = r_{16}$ & $1.2\pm 1.3$ ($1.3\pm1.2$) & $[0.10,4.2]$ ($[0.13,3.9]$) \\
  $x_1 = \Gamma/r_{16}$ & $194\pm 13$ ($215\pm 3.4$) & $[173,214]$ ($[210,220]$) \\
  $x_2 = N_{\pm,57}/r_{16}$ & $3.0\pm 0.19$ ($2.8\pm 0.17$) & $[2.7,3.3]$ ($[2.5,3.1]$) \\
  \hline
  $\Gamma$ &  $240\pm 240$ ($270\pm260$ ) & $[20,800]$ ($[27,830]$) \\
  $N_{\pm,57}$ & $3.7\pm 3.9$ ($3.6\pm3.4$) & $[0.30,12]$ ($[0.33,11]$) \\
  $t_0/\mathrm{s}$ & $234 \pm  17$ ($238 \pm 10$) & $[217,270]$ ($[229,260]$) \\
  \end{tabular}\tablefoot{Each line refers to a model parameter (see Appendix \ref{sec:HLE_fitting}) or a quantity derived from a combination of model parameters, the two cases being separated with a horizontal line. For each line, we report the mean, standard deviation, and symmetric 90\% credible range (comprised between the 5$^\mathrm{th}$ and 95$^\mathrm{th}$ percentiles) of the posterior samples obtained by fitting the line properties. Results in parentheses are for the fit that includes the data in the BTI.}
 \label{tab:HLE_fit_results}
\end{table}

The interpretation of the line as HLE of $e^+e^-$ pair annihilation implies some additional constraints on $t_0$, $r$ and $\Gamma$, which follow from two requirements: (i) the start time $t_0$ must be earlier than the first observation of the line, and (ii) the HLE-predicted evolution must reproduce that inferred from the observations. These conditions are somewhat delicate, because of saturation of the GBM detectors during the `Bad Time Interval' \citep[BTI, from 219 to 277 s, as given in][and shown here with gray shading in Fig.\ \ref{fig:lc_and_line_props}]{Lesage2023} and because different analyses do not fully agree on the inferred properties of the line, as shown by the presence of incompatible error bars in Fig.\ \ref{fig:lc_and_line_props}. A possible reason for the discrepant results at times $t\gtrsim 320$ s could reside in the difficulty in modelling the continuum: as shown by the analysis reported by \citetalias{Ravasio2024}, also confirmed by \citet{Zhang2024_observation} and \citet{Axelsson2024}, at these times an additional emission component is present, on top of the usual prompt emission continuum, which is probably related to the emergence of the afterglow. Different choices in the parametrization of the emission components and other technical details could have led to inconsistent results.

With these caveats in mind, we fitted the simple HLE physical model (Eqs.\ \ref{eq:L_HLE} and \ref{eq:hnu_HLE}) to two sub-samples of the data points shown in Fig.\ \ref{fig:lc_and_line_props}, selected as follows. We ignored data beyond $t=320$ s and we did not take into account the results from \citet[][because of the discontinuous $h\nu_\mathrm{line}$ implied there, which is in contradiction with the assessment from the other analyses]{Axelsson2024} and from \citet[][because of the low significance]{Frederiks2023}. Both sub-samples hence comprise datapoints from \citetalias{Ravasio2024} and \citet{Zhang2024_observation}: in one case, we conservatively excluded the line properties inferred by \citet{Zhang2024_observation} using data earlier than $t=275$ s (i.e.\ BTI data), while in the other case we included all their datapoints. Appendix \ref{sec:HLE_fitting} provides technical details about the procedure used. The results are summarized in Table \ref{tab:HLE_fit_results}, where number in parentheses refer to the analysis that includes BTI data. In Fig.\ \ref{fig:HLE_fit_results} we also show with cyan (magenta) lines 100 posterior samples of the model line luminosity and photon energy evolution obtained from the fit excluding (including) the data within the BTI. The associated posterior probability density of $t_0$ is shown with the corresponding color in the bottom panel.

Within the allowed range, as expected based on the analytical reasoning in the previous section, the preferred values of $\Gamma$ and $N_\pm$ are highly correlated with $r$ and can be summarized as $\Gamma = (194\pm 13)r_{16}$ and $N_\pm= (3.0\pm 0.19)\times 10^{57} r_{16}$ when excluding BTI data, or $\Gamma = (215\pm 3.4)r_{16}$ and $N_\pm= (2.8\pm 0.17)\times 10^{57}r_{16}$ when including it. These results, while formally in a slight tension, are qualitatively consistent with those obtained in the previous section.
The model fitting produces an additional, informative constraint on the arrival time of the first photon, $t_0 = 234\pm 17\,\mathrm{s}$ if BTI data are excluded, or $t_0 = 238\pm 10\,\mathrm{s}$ if  included. In both cases, this is strikingly close to the initial sharp rise in the LHAASO light curve, as shown in Fig.\ \ref{fig:HLE_fit_results}.

These results show that the HLE interpretation of the line evolution, within the $e^+e^-$ pair annihilation scenario, requires the spherical shell where the annihilation took place to be quite rapidly expanding, geometrically as thin as an external shock (or thinner), and to contain a large number of pairs. Its radius must have been somewhere between about $10^{15}$ cm up to around $4\times 10^{16}$ cm, and the arrival time of the first photon must have been within the brightest peak of the prompt emission and close to the sharp rising phase at the beginning of the LHAASO afterglow.

\revone{We note that the constraints obtained here are different from those derived by \citet{Yi2024} based on optical depth arguments. One difference is that their assumptions break down in the case of HLE: in deriving the number of annihilating pairs, they assume that the comoving emitting volume $V^\prime$ is the same at all times; in the case of HLE, on the other hand, only an annulus of constant latitude contributes to the emission at each fixed time.}

\subsection{Pair multiplicity}

In the range of radii derived above, the external shock that arises from the interaction of the GRB ejecta with the external medium is expected to become important. Let us assume an external medium whose composition is described by an electron fraction $Y_\mathrm{e}$ ($Y_\mathrm{e}=1$ for pure hydrogen, $Y_\mathrm{e}\sim 0.5$ for other elements), with a number density profile $n_\mathrm{ext}(r)=A_\star(r/r_\star)^{-k}$, where $A_\star = 1\,A_{\star,0}\,\mathrm{cm^{-3}}$ is the number density at a reference radius $r_\star\equiv 5.5\times 10^{17}\,\mathrm{cm}$. This definition ensures that our reference density $A_{\star}=1\,\mathrm{cm^{-3}}$ matches that of a typical interstellar medium (ISM) if $k=0$, and that of a typical Wolf-Rayet stellar wind \citep{Chevalier2000} if $k=2$. If $k=0$, then $n_\mathrm{ext}$ is equal to $A_{\star}$ at all radii, and we will use the former symbol for clarity. If the pairs are created downstream of an external shock, at a radius $r$ this would have swept a number of electrons $N_\mathrm{e}=4\pi Y_\mathrm{e} A_\star r^{3-k}r_\star^k/(3-k)$. The pair multiplicity, that is, the number of leptons per external electron that need to be produced in the HLE scenario (focusing on the most relevant cases $k=0$ and $k=2$) is
\begin{equation}
\begin{split}
 & Z_{\pm,\mathrm{HLE}} = \frac{N_\mathrm{e}+2 N_\pm}{N_\mathrm{e}}\approx\\
 & 1+\left\lbrace\begin{array}{lr}
8.5\times 10^{6} Y_\mathrm{e}^{-1} r_{16}^{-2}n_{\mathrm{ext},2}^{-1}L_\mathrm{line,50}\left(\frac{h\nu_\mathrm{line}}{12.6\,\mathrm{MeV}}\right)^{-3}\left(\frac{1+z}{1.15}\right)^{-3}, & k=0;\\
9.4\times 10^{4}Y_\mathrm{e}^{-1} A_{\star,0}^{-1}L_\mathrm{line,50}\left(\frac{h\nu_\mathrm{line}}{12.6\,\mathrm{MeV}}\right)^{-3}\left(\frac{1+z}{1.15}\right)^{-3},  & k=2.\\
\end{array}\right.
\end{split}
\label{eq:Zpm}
\end{equation}
This very high multiplicity requires a scenario where an efficient mechanism of pair production can operate at the required radii. In the following, we set out to construct such a scenario.

\section{The scenario}
\label{sec:Scenario}
\begin{figure*}
 \includegraphics[width=\textwidth]{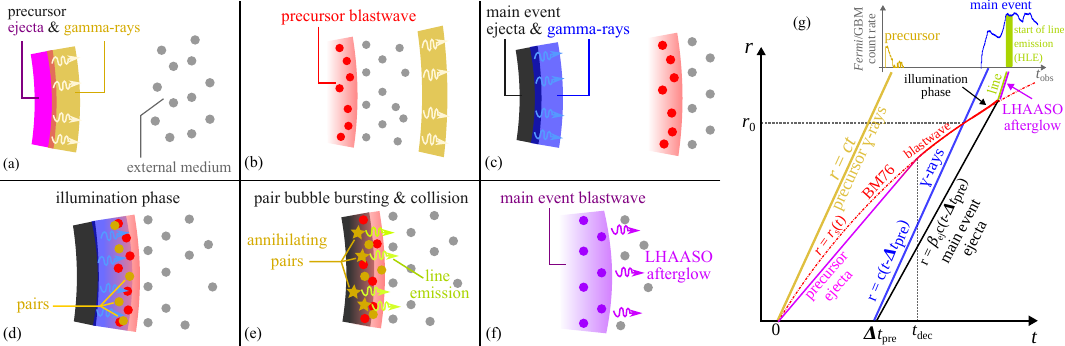}
\caption{Sketch of the proposed scenario (not to scale). After emitting their gamma-rays (a), the precursor ejecta expand into the external medium and drive a blastwave (b). The main event ejecta emit their gamma-rays (c) and initially expand in the cavity left by the precursor. The main event gamma-ray front illuminates the precursor blastwave (d) leading to copious pair creation and radiative acceleration. Soon after, the main event ejecta collide with the pair-enriched blastwave (e). The mass-annihilation of the pairs (`pair bubble bursting') happens somewhere between the end of (d) and the end of (e). When the main-event-ejecta-driven shock reaches the external medium, it leads to a much more energetic blastwave (f), whose emission (including the component seen by LHAASO) rises steeply before settling to the usual, self-similar evolution. Panel (g) shows the evolution of the radii of the components described in the other panels as a function of the `lab-frame' time $t$ in the progenitor rest frame. The inset shows the \textit{Fermi}/GBM light curve, highlighting the connection between $t$ and the observer-frame time $t_\mathrm{obs}$.}
 \label{fig:scenario}
\end{figure*}

A peculiar feature of \GRB\ is the presence of a `precursor': the emission starts with a short burst lasting a few seconds, with an energy  $E_{\gamma,\mathrm{pre}}\sim 10^{51}\,\mathrm{erg}$ radiated in gamma-rays, followed by a quiescent period of rest-frame duration $\Delta t_\mathrm{pre}\sim 160\,\Delta t_\mathrm{pre,2.2}\,\mathrm{s}$, after which the main emission (which delivers a $10^4$-fold larger energy -- \citealt{Frederiks2023}) starts off (Fig.\ \ref{fig:lc_and_line_props}, panel a). The precursor, which constitutes by itself a relatively powerful gamma-ray burst, must be associated with a relativistic outflow.
The expansion of the ejecta of such an outflow into the external medium drives a relativistic blastwave \citep[][BM76 hereafter]{Blandford1976}.
When the main event photons illuminate this blastwave with their extremely large luminosity and hard spectrum, they can modify the composition and dynamics by interacting with leptons in the shocked material. We envisage a scenario where this `illumination phase' leads to copious pair production, through a similar process as that proposed by \citet{Madau2000} and \citet{Thompson2000}, but with the precursor blastwave as a target, instead of a cold external medium. In the following we show that, under some conditions, the illumination phase ends with (or is followed by) the `pair bubble bursting', that is, mass-annihilation of the created pairs. This leads to the appearance of the line, which is seen to evolve due to the HLE effect. In this scenario, the alignment of the start time $t_0$ of the HLE with the onset of the LHAASO afterglow has a natural explanation: before colliding with the precursor blastwave, the main event ejecta expanded in vacuum, hence they did not contribute to the afterglow. With the collision, the energy in the external shock increases by orders of magnitude, causing a sudden, sharp rise in the luminosity. The collision happens right after the illumination phase, therefore it is almost contemporaneous to the pair bubble bursting.

In Fig. \ref{fig:scenario} we sketch the main phases of the proposed scenario, along with a depiction of the time evolution of the radii of the precursor ejecta and gamma-rays, the precursor blastwave, and of the main event ejecta and gamma-rays. In the following we describe in detail the initial conditions of the illumination phase and the requirements that the scenario imposes on the properties of the external medium and of the precursor ejecta.

\subsection{Precursor blastwave radius and Lorentz factor at the start of the illumination phase}

Let us indicate with $E_\mathrm{pre}=10^{52}\,E_\mathrm{pre,52}\,\mathrm{erg}$ the energy in the precursor ejecta and with $\Gamma_\mathrm{ej,pre}$ their bulk Lorentz factor. The associated `deceleration' time (measured in the rest frame of the progenitor) is
\begin{equation}
 t_\mathrm{dec}\approx \left(\frac{(3-k)E_\mathrm{pre}}{4\pi A_\star r_\star^k m_\mathrm{p}c^2 \Gamma_\mathrm{ej,pre}^{2} c^{3-k}}\right)^{1/(3-k)},
\end{equation}
where $m_\mathrm{p}$ is the proton rest mass.
The deceleration, which corresponds to the time when the most of the ejecta energy is deposited into a relativistic blastwave that expands into the external medium, is seen to happen at a time
$t_\mathrm{dec,obs}\approx (1+z)t_\mathrm{dec}/\Gamma_\mathrm{ej,pre}^2$
by a distant observer.

Let us focus on panel (g) of Fig. \ref{fig:scenario}. At $t\ll t_\mathrm{dec}$ the precursor ejecta (purple) propagate at constant speed and produce a photon pulse (the precursor emission -- yellow).  At a time $t>t_\mathrm{dec}$, the ejecta deposited all their energy into a blastwave (red) whose leading shock propagates with a Lorentz factor (\citetalias{Blandford1976})
\begin{equation}
 \Gamma_\mathrm{s}(t) \approx \left[\frac{(17-4k) E_\mathrm{pre}}{8 \pi A_\star r_\star^k m_\mathrm{p} c^2 (c t)^{3-k}}\right]^{1/2},
\label{eq:bw_LF}
 \end{equation}
with a radius evolving as (dot-dashed red line)
\begin{equation}
 r_\mathrm{s}(t) \approx \left(1-\frac{1}{2(4-k)\Gamma_\mathrm{s}^2(t)}\right)ct.
 \label{eq:rsBM76}
\end{equation}
The main event ejecta (black) initially expand at constant speed into the cavity left behind by the precursor blastwave, and produce their gamma-ray emission (blue). The latter illuminates the blastwave when its radius is $r_0$ (see below). In our scenario, we assume this to happen later than the deceleration time. This corresponds to the condition $t_\mathrm{dec,obs}<\Delta t_\mathrm{pre}$, which implies a precursor ejecta bulk Lorentz factor $\Gamma_\mathrm{ej,pre}\gtrsim 111\,E_\mathrm{pre,52}^{1/8}n_{\mathrm{ext},0}^{-1/8}\Delta t_\mathrm{pre,2.2}^{-3/8}$ for $k=0$ or $\Gamma_\mathrm{ej,pre}\gtrsim 26\,E_\mathrm{pre,52}^{1/4}A_{\star,0}^{-1/4}\Delta t_\mathrm{pre,2.2}^{-1/4}$ for $k=2$.

In the precursor blastwave, most of the shocked external medium material is confined within a thin shell of width $\Delta r_\mathrm{b}\sim r_\mathrm{s}/\Gamma_\mathrm{s}^2$ behind the shock. Moreover, when the outer edge of the gamma-ray front is located at a radius $r$, the main event ejecta lag only shortly behind, at a distance $\Delta r_\mathrm{\gamma,ej} \approx r/2\Gamma_\mathrm{ej}^2 \lesssim \Delta r_\mathrm{b}$, where $\Gamma_\mathrm{ej}$ is the main event ejecta bulk Lorentz factor. Hence, the interaction between the main event gamma-rays and the blastwave, before the main event ejecta catch up, happens entirely around the radius $r_0$ at which the gamma-ray front starts illuminating the blastwave shock. To compute this radius, let us consider a light pulse whose arrival time to the observer is delayed by $\Delta t_\mathrm{pre}$ with respect to the precursor emission. The radius of such pulse (blue line in panel g of Fig.\ \ref{fig:scenario}) evolves simply as $r_\gamma = c(t-\Delta t_{\mathrm{pre}})$. The gamma-ray front reaches the shock at a radius
\begin{equation}
 r_0 \approx \left[\frac{(17-4k)(4-k)E_\mathrm{pre}c\Delta t_\mathrm{pre}}{4\pi A_\star r_\star^k m_\mathrm{p}c^2}\right]^{1/(4-k)}.
\label{eq:r0}
 \end{equation}
For the two external medium profiles of interest, we have
\begin{equation}
 r_0 \approx\left\lbrace
 \begin{array}{lr}
   3.6\times 10^{16}\,E_\mathrm{pre,52}^{1/4}n_{\mathrm{ext},2}^{-1/4}\Delta t_\mathrm{pre,2.2}^{1/4}\,\mathrm{cm}, & k=0;\\
  1.3\times 10^{16}\,E_\mathrm{pre,52}^{1/2}A_{\star,0}^{-1/2}\Delta t_\mathrm{pre,2.2}^{1/2}\,\mathrm{cm}, & k=2.
 \end{array}\right.
\end{equation}
In the $k=0$ case, keeping the precursor ejecta kinetic energy fixed at the reference value, the external medium density must be at least $n_\mathrm{ext,2}\gtrsim 0.55 E_\mathrm{pre,52}^{-1}$ for $r_0$ to be in the range required by the HLE scenario (see Table \ref{tab:HLE_fit_results}), which explains why we take $n_\mathrm{ext}=10^2\,\mathrm{cm}^{-3}$ as reference. An even larger $n_\mathrm{ext}$ in this case would also help in alleviating the extremely stringent requirement on the pair multiplicity (Eq.\ \ref{eq:Zpm}). In the $k=2$ case, the density cannot be higher than $A_{\star,0}\approx 160 E_\mathrm{pre,52}^{-1}$, otherwise $r_0$ would become smaller than the minimum allowed value in the HLE scenario, $\sim 10^{15}\,\mathrm{cm}$.

From Eq.~\ref{eq:bw_LF}, the corresponding Lorentz factor of the material right behind the blastwave shock, $\Gamma_0\sim \Gamma_\mathrm{s}(r_0)/\sqrt{2}$, is
\begin{equation}
 \Gamma_0\approx\left\lbrace
 \begin{array}{lr}
  22 \,E_\mathrm{pre,52}^{1/8}n_{\mathrm{ext},2}^{-1/8}\Delta t_\mathrm{pre,2.2}^{-3/8}, & k=0;\\
  18 \,E_\mathrm{pre,52}^{1/4}A_{\star,0}^{-1/4}\Delta t_\mathrm{pre,2.2}^{-1/4}, & k=2.
  \label{eq:Gamma0}
 \end{array}\right.
\end{equation}

By substituting $r_0$ back into the expressions for the Lorentz factor and pair multiplicity required in the HLE scenario, Eqs.\ \ref{eq:GammaHLE} and \ref{eq:Zpm}, we can see how the HLE requirements depend on the blastwave parameters. For the Lorentz factor, we obtain
\begin{equation}
 \Gamma_\mathrm{HLE} \approx \left\lbrace\begin{array}{lr}
  815\,E_\mathrm{pre,52}^{1/4}n_{\mathrm{ext},2}^{-1/4}\Delta t_\mathrm{pre,2.2}^{1/4}\left(\frac{t_\mathrm{line}-t_0}{60\,\mathrm{s}}\right)^{-1}\left(\frac{h\nu_\mathrm{line}}{12.6\,\mathrm{MeV}}\right)^{-1} & k=0;\\
  276\,E_\mathrm{pre,52}^{1/2}A_{\star,0}^{-1/2}\Delta t_\mathrm{pre,2.2}^{1/2}\left(\frac{t_\mathrm{line}-t_0}{60\,\mathrm{s}}\right)^{-1}\left(\frac{h\nu_\mathrm{line}}{12.6\,\mathrm{MeV}}\right)^{-1} & k=2,
  \end{array}\right.
\end{equation}
or
\begin{equation}
 \frac{\Gamma_\mathrm{HLE}}{\Gamma_0} \approx \left\lbrace\begin{array}{lr}
  37\,E_\mathrm{pre,52}^{1/8}n_{\mathrm{ext},2}^{-1/8}\Delta t_\mathrm{pre,2.2}^{5/8}\left(\frac{t_\mathrm{line}-t_0}{60\,\mathrm{s}}\right)^{-1}\left(\frac{h\nu_\mathrm{line}}{12.6\,\mathrm{MeV}}\right)^{-1} & k=0;\\
  15\,E_\mathrm{pre,52}^{1/4}A_{\star,0}^{-1/4}\Delta t_\mathrm{pre,2.2}^{3/4}\left(\frac{t_\mathrm{line}-t_0}{60\,\mathrm{s}}\right)^{-1}\left(\frac{h\nu_\mathrm{line}}{12.6\,\mathrm{MeV}}\right)^{-1} & k=2.
  \end{array}\right.
\end{equation}
This shows that, along with pair-enrichment, some acceleration mechanism is needed in order to meet the required Lorentz factor. In Section \ref{sec:acceleration} and in Appendix \ref{sec:collision_phase} we discuss two possible such mechanisms.

In the $k=0$ case, we can express the pair multiplicity required in the HLE scenario, Eq.\ \ref{eq:Zpm}, as a function of the blastwave parameters, namely
\begin{equation}
\begin{split}
 & Z_\mathrm{\pm,HLE} -1\approx\\
 & 6.5\times 10^5\,Y_\mathrm{e}^{-1} E_\mathrm{pre,52}^{-1/2}n_{\mathrm{ext},2}^{-1/2}\Delta t_\mathrm{pre,2.2}^{-1/2}L_\mathrm{line,50}\left(\frac{h\nu_\mathrm{line}}{12.6\,\mathrm{MeV}}\right)^{-3}\left(\frac{1+z}{1.15}\right)^{-3}.
\end{split}
\end{equation}
The $k=2$ case remains the same as in Eq.\ \ref{eq:Zpm}.

\subsection{Energy in gamma-ray front and ejecta Lorentz factor}
\label{sec:Egf}

Let us indicate the energy in gamma-ray photons that flow through the blastwave during the illumination phase as $E_\mathrm{\gamma f}$, and the associated average luminosity with $L$. In our scenario, the illumination phase ends at the time when the LHAASO light curve rises, that is around $t_\mathrm{obs}\sim 226$ s. Taking the results of the spectral analysis of \citet{Frederiks2023} in the time interval $t_\mathrm{obs}\in [180,233]\,\mathrm{s}$ as reference, we obtained $L \approx 4\times 10^{52}\,\mathrm{erg\,s^{-1}}$ and $E_\mathrm{\gamma f}\approx 1.6\times 10^{54}\,\mathrm{erg}$. The fraction of this energy that needs to be converted into pairs to fulfil the requirement from Eq.\ \ref{eq:Npm} is
\begin{equation}
 \frac{2 \Gamma_\mathrm{HLE} N_\pm m_\mathrm{e}c^2}{E_\mathrm{\gamma f}} \approx 0.4\,r_{16}^2L_\mathrm{line,50}\left(\frac{h\nu_\mathrm{line}}{12.6\,\mathrm{MeV}}\right)^{-4}\left(\frac{1+z}{1.15}\right)^{-3}\left(\frac{t_\mathrm{line-t_0}}{60\,\mathrm{s}}\right)^{-1}.
\end{equation}
This shows that the main event gamma-ray front contains in principle sufficient energy to produce the required pair loading, in presence of an efficient enough pair production mechanism.

As noted above, the radial extent of the gamma-ray front is
$\Delta r_\mathrm{\gamma,ej} \sim r_0/2 \Gamma_\mathrm{ej}^2 $, and we have $E_\mathrm{\gamma f}/L \sim \Delta r_\mathrm{\gamma,ej}/c$.   Therefore, the bulk Lorentz factor of the main event ejecta must have been
\begin{equation}
 \Gamma_\mathrm{ej}\approx 65\,r_{0,16}^{1/2} \left(\frac{E_\mathrm{\gamma f}/L}{40\,\mathrm{s}}\right)^{-1/2}
\end{equation}
in order for the assumptions behind our scenario to hold. At first sight, this low bulk Lorentz factor may seem in contradiction with that required by compactness arguments, $\Gamma_\mathrm{ej,0}\gtrsim 300-1000$ \citep{Lesage2023,Gao2023}. However, we point out that (i) the collision involves at first the outermost part of the ejecta, which are associated with the first pulse of the main event (that extends approximately from 180 s to 210 s, Fig.\ \ref{fig:lc_and_line_props}). This pulse is less luminous and has a softer spectrum with respect to the most luminous peak, on which the \citet{Lesage2023} analysis is based; (ii) $\Gamma_\mathrm{ej}$ here refers to the bulk Lorentz factor of the ejecta long after the prompt emission, while the compactness limit applies to the bulk Lorentz factor $\Gamma_\mathrm{ej,0}$ of the ejecta in the act of producing such emission. In the case of a high prompt emission efficiency $\eta_\gamma$ (which represents the fraction of the outflow energy that is radiated in gamma-rays), the radiation carries away sufficient energy and momentum to slow down significantly the ejecta. Indeed, energy and rest-mass conservation require $\Gamma_\mathrm{ej}=(1-\eta_\gamma)\Gamma_\mathrm{ej,0}$. The limits from the compactness argument therefore do not conflict directly with our scenario, provided that a very high efficiency $\eta_\gamma = 1-\Gamma_\mathrm{ej}/\Gamma_\mathrm{ej,0} \sim 0.8-0.95$ could be realized in the main event gamma-ray emission, assuming $300\lesssim \Gamma_\mathrm{ej,0}\lesssim 1000$ and $r_{0,16}=1$. Conversely, a less extreme efficiency $\eta_{\gamma}=0.5$, would lead to $\Gamma_\mathrm{ej,0}\approx 130\,r_{0,16}^{1/2}$, which could be reasonable for the ejecta associated to the first pulse of the main event.

For what concerns the Lorentz factor estimates based on the assumption that the LHAASO peak corresponds to the deceleration time \citep{LHAASO2023,Lesage2023}, we note that these neglect the presence of a precursor and its effect on the external medium, but rather arbitrarily shift the explosion time in order to match the initial rise power law slope expected in a scenario where the external medium has a uniform density. These estimates therefore cannot be compared to ours.

\subsection{Requirements set by acceleration to the Lorentz factor implied by the HLE scenario}
\label{sec:acceleration}

Two acceleration mechanisms operate on the blastwave after the start of the illumination phase: the momentum deposition by main event photons (radiative acceleration) and the subsequent hydrodynamical acceleration when the main event ejecta collide with the blastwave. As discussed later, we found that, for our source, only the former is relevant, hence we defer the discussion of the latter to Appendix \ref{sec:collision_phase}.

As discussed in Sect.\ \ref{sec:HLErequirements}, if the pair multiplicity reaches the required level, then the blastwave becomes a pair-dominated shell  (Eq.\ \ref{eq:Zpm}). Therefore, most of the rest mass is in pairs, $M_\mathrm{shell}\sim 2 N_\pm m_\mathrm{e}$. If the shell is accelerated to the bulk Lorentz factor given in Eq.\ \ref{eq:GammaHLE}, its kinetic energy is roughly
\begin{equation}
\begin{split}
&\Gamma_\mathrm{HLE} M_\mathrm{shell}c^2\sim 2\Gamma_\mathrm{HLE} N_\pm m_\mathrm{e}c^2 \approx\\
&\approx 6.7 \times 10^{53}\,r_{16}^2L_\mathrm{line,50}\left(\frac{h\nu_\mathrm{line}}{12.6\,\mathrm{MeV}}\right)^{-4}\left(\frac{1+z}{1.15}\right)^{-4}\left(\frac{t_\mathrm{line}-t_0}{60\,\mathrm{s}}\right)^{-1}\,\mathrm{erg}.
\end{split}
\end{equation}
For $r_{16} \lesssim 1.5$, we have $\Gamma_\mathrm{HLE}M_\mathrm{shell}c^2 \lesssim E_{\gamma\mathrm{f}}$, therefore radiative acceleration could be efficient enough to meet the HLE requirement during the illumination phase.

\subsection{Requirements set by pair cooling and annihilation}

Pairs can cool due to inverse Compton (IC) scattering of main event photons, which, after the acceleration of the blastwave, happens mostly in the Thomson regime. The IC cooling time scale is $t_\mathrm{cool}^\prime\sim 3\pi m_\mathrm{e}c^2 r^2 \Gamma^2 / \sigma_\mathrm{T}L$, while the pair annihilation time (for cold pairs) is $t^\prime_\mathrm{ann}\sim 8/3\sigma_\mathrm{T}c n_\pm^\prime$. For the pair annihilation to produce a narrow line, the cooling time scale must be shorter than the annihilation time scale. Assuming again $V^\prime = 4\pi \xi r^3/\Gamma$ (see Section \ref{sec:HLErequirements}) and substituting $N_\pm$ from Eq.\ \ref{eq:Npm} and $\Gamma$ from Eq.\ \ref{eq:GammaHLE}, we obtain
\begin{equation}
 \frac{t^\prime_\mathrm{ann}}{t^\prime_\mathrm{cool}}\approx 0.3\,r_{16}^{-3}\xi_{-1}L_{52.6} L_\mathrm{line,50}^{-1} \left(\frac{h\nu_\mathrm{line}}{12.6\,\mathrm{MeV}}\right)^6 \left(\frac{1+z}{1.15}\right)^3 \left(\frac{t_\mathrm{line}-t_0}{60\,\mathrm{s}}\right)^3.
\end{equation}
Therefore, pairs cool faster than they annihilate as long as
\begin{equation}
 r_{16}\lesssim 0.7\,\xi_{-1}^{1/3}L_{52.6}^{1/3} L_\mathrm{line,50}^{-1/3} \left(\frac{h\nu_\mathrm{line}}{12.6\,\mathrm{MeV}}\right)^2 \left(\frac{1+z}{1.15}\right) \left(\frac{t_\mathrm{line}-t_0}{60\,\mathrm{s}}\right).
\end{equation}

These calculations show that, within a sizable fraction of the parameter space allowed by the HLE scenario, the pairs are expected to cool efficiently (and hence produce a narrow annihilation line) and to annihilate fast enough for the HLE effect to dominate the observed line evolution.

\section{The illumination phase: pair enrichment and radiative acceleration} \label{sec:pair_enrichment}

\subsection{Non-thermal electrons in the blastwave}\label{sec:gamma_distributions}
In order to work out the effect of the main event photons on the blastwave, we need to describe the leptonic content of the latter.
We assume electrons from the external medium to be accelerated at the blastwave shock, with an injection rate  distributed as a power law in the electron Lorentz factor, $\mathrm{d}\dot n_\mathrm{e,inj}^\prime/\mathrm{d}\gamma \propto \gamma^{-p}$, and we fix the power law index at $p=2.2$ based on the results of particle-in-cell simulations of relativistic shocks in weakly magnetized media \citep{Sironi2013}. After leaving the acceleration region, electrons keep interacting with the shock-generated, turbulent magnetic field, producing synchrotron emission. Electrons can also interact with synchrotron photons, producing synchrotron-self Compton (SSC) emission. These emission processes affect the electron energy distribution across the blastwave.

We adopt here a simplified description of the blastwave structure as a homogeneous shell whose density is given by the ultra-relativistic Rankine-Hugoniot condition $n \sim 4\Gamma_0\Gamma_0 n_\mathrm{ext}(r_0)$. Conservation of the external medium rest mass then implies an effective shell thickness $\Delta r_\mathrm{0} = r_0/4(3-k)\Gamma_0^2$.

Assuming the acceleration process to affect a fraction $\xi_\mathrm{N}\sim 1$ of the external medium electrons, and defining $\epsilon_\mathrm{e}=0.1\,\epsilon_\mathrm{e,-1}$ as the fraction of the shock downstream internal energy that is shared by the accelerated electrons, we have that the minimum Lorentz factor of injected electrons is $\gamma_\mathrm{m}\sim(p-2)\epsilon_\mathrm{e}\Gamma_0 m_\mathrm{p}/(p-1)m_\mathrm{e}$, which gives
\begin{equation}
 \gamma_\mathrm{m}\approx\left\lbrace\begin{array}{lr}
670\,\epsilon_\mathrm{e,-1}E_\mathrm{pre,52}^{1/8}n_\mathrm{ext,2}^{-1/8}\Delta t_\mathrm{pre,2.2}^{-3/8}, & k=0;\\
550 \,\epsilon_\mathrm{e,-1}E_\mathrm{pre,52}^{1/4}A_{\star,0}^{-1/4}\Delta t_\mathrm{pre,2.2}^{-1/4}, & k=2.
                                     \end{array}\right.
\label{eq:gm}
\end{equation}

Most of the electrons contained in the shell were accelerated during the last dynamical time $t^\prime_\mathrm{dyn}=r_0/(\Gamma_0 c)$. Over that time, electrons accelerated to an initial Lorentz factor larger than $\gamma_\mathrm{c}$ had time to lose most of their energy by synchrotron and synchrotron-self-Compton radiation, with
\begin{equation}
 \gamma_\mathrm{c} \sim \frac{6\pi m_\mathrm{e} c^2 \Gamma_0}{\sigma_\mathrm{T}r_0 B^{\prime 2} (1+Y^\prime)}\equiv \frac{\gamma_\mathrm{c,syn}}{1+Y^\prime},
\end{equation}
where $\sigma_\mathrm{T}$ is the Thomson cross section, $B^\prime$ is the average strength of the turbulent magnetic field in the shock downstream, $Y^\prime$ is the `Compton parameter', that is, the ratio of the SSC to the synchrotron losses for electrons with $\gamma=\gamma_\mathrm{c}$. Assuming the turbulent magnetic field to share a fraction $\epsilon_\mathrm{B}=10^{-3}\,\epsilon_\mathrm{B,-3}$ of the shock downstream internal energy, we have $B^\prime=\sqrt{32\pi\epsilon_\mathrm{B}\Gamma_0^2 n_\mathrm{ext}(r_0) m_\mathrm{p}c^2}$. For our early time of interest, we find that SSC cooling dominates over synchrotron, with \citep[see Eq.\ 35 in][]{Panaitescu2000}
\begin{equation}
1+Y^\prime = \frac{1}{2}\left[-1+\sqrt{\frac{(5-k)\epsilon_\mathrm{e}}{2(3-k)\epsilon_\mathrm{B}}+1}\right]\approx 
\left\lbrace\begin{array}{lr}
  4.1 \epsilon_\mathrm{e,-1}^{1/2}\epsilon_\mathrm{B,-3}^{-1/2}
   &  k=0;\\
5.6 \epsilon_\mathrm{e,-1}^{1/2}\epsilon_\mathrm{B,-3}^{-1/2}
     & k=2,
\end{array}\right.
\end{equation}
which then gives
\begin{equation}
 \gamma_\mathrm{c} \sim \left\lbrace\begin{array}{lr}
 480\,\epsilon_\mathrm{e,-1}^{-1/2}\epsilon_\mathrm{B,-3}^{-1/2} n_\mathrm{ext,2}^{-5/8} E_\mathrm{pre,52}^{-3/8}\Delta t_\mathrm{pre,2.2}^{1/8}, &  k=0;\\
 62\,\epsilon_\mathrm{e,-1}^{-1/2}\epsilon_\mathrm{B,-3}^{-1/2} A_\mathrm{\star,0}^{-5/4} E_\mathrm{pre,52}^{1/4}\Delta t_\mathrm{pre,2.2}^{3/4}, & k=2.
\end{array}\right.
\label{eq:gc}
 \end{equation}

As a result of acceleration and cooling, the Lorentz factor distribution of the electrons in the blastwave approximately follows a broken power law, which can be summarized as
\begin{equation}
 \frac{\mathrm{d}n_\mathrm{e}^\prime}{\mathrm{d}\gamma} \sim \frac{n_\mathrm{e}^\prime}{\gamma_\mathrm{c}} \left\lbrace\begin{array}{lr}
 \left(\frac{\gamma}{\gamma_\mathrm{c}}\right)^{-2} & \gamma_\mathrm{c}\leq\gamma < \gamma_\mathrm{m}  \\
 \left(\frac{\gamma_\mathrm{m}}{\gamma_\mathrm{c}}\right)^{-2}\left(\frac{\gamma}{\gamma_\mathrm{m}}\right)^{-p-1} & \gamma_\mathrm{m}\leq \gamma \\
 0 & \mathrm{otherwise}
 \end{array}\right.,
 \label{eq:dne_dg}
\end{equation}
which shows that most electrons in the blastwave have a Lorentz factor close to $\gamma_\mathrm{c}$.

\subsection{Incident main event spectrum and inverse Compton cooling of the blastwave original electrons}
\label{sec:incident_spec}

As soon as the blastwave is illuminated by the main event photons, these can be subject to inverse Compton scattering off the blastwave electrons. Let $\varepsilon$ be a main event photon energy in units of the electron rest mass energy, and let $L_\varepsilon$ be the specific luminosity of the main event emission per unit of $\varepsilon$. The spectrum of the latter can be described as
\begin{equation}
L_\varepsilon = \frac{L}{\varepsilon_\mathrm{peak}}f_\mathrm{main}\left(\frac{\varepsilon}{\varepsilon_\mathrm{peak}}\right),
\end{equation}
where we defined the main event spectral shape following the empirical `Band' function \citep{Band1993}
\begin{equation}
 f_\mathrm{main}(x)\propto \left\lbrace\begin{array}{lr}
x^{1-\alpha_1}\mathrm{e}^{-(2-\alpha_1)x} & x\leq \frac{(\alpha_2-\alpha_1)}{(2-\alpha_1)} \\
x^{1-\alpha_2}\left(\frac{\alpha_2-\alpha_1}{2-\alpha_1}\right)^{\alpha_2-\alpha_1}\mathrm{e}^{\alpha_1-\alpha_2} & \frac{(\alpha_2-\alpha_1)}{(2-\alpha_1)}<x\leq x_\mathrm{cut}  \\
0 & x>x_\mathrm{cut}
\end{array}\right.,
\end{equation}
with $x=\varepsilon/\varepsilon_\mathrm{peak}$.

As per the spectral analysis reported in \citet{Frederiks2023}, the average spectrum of the main event has $\alpha_1\approx 0.89$, $\alpha_2\approx 2.21$ and $E_\mathrm{peak}\approx 2660\,\mathrm{keV}$, which translates into $\varepsilon_\mathrm{peak}=(1+z)E_\mathrm{peak}/m_\mathrm{e}c^2\approx 6$. As noted in Section \ref{sec:Egf}, during our period of interest, using the results from the analysis in the same paper, we have $L\approx 4\times 10^{52}\,\mathrm{erg\,s^{-1}}$ (integrated in the 20 keV -- 10 MeV band). The normalization of $f_\mathrm{main}$ is set so that
\begin{equation}
\int_{(1+z)20\,\mathrm{keV}/\varepsilon_\mathrm{peak}}^{(1+z)10\,\mathrm{MeV}/\varepsilon_\mathrm{peak}} f_\mathrm{main}(x)\mathrm{d}x=1.
\end{equation}
In the spectral shape, we introduced a high-energy cut-off at an observed photon energy $E_\mathrm{cut}=x_\mathrm{cut}\varepsilon_\mathrm{peak}m_\mathrm{e}c^2/(1+z)$ (discussed later).

The typical electron in the blastwave sees the incident spectral luminosity peak at $\varepsilon^{\prime\prime}\sim \gamma_\mathrm{c}\varepsilon_\mathrm{peak}/\Gamma_0$. The IC scattering of photons at that peak is in the Thomson regime provided that $\varepsilon^{\prime\prime}\lesssim 1$, which is verified as long as
\begin{equation}
\begin{array}{lr}
 n_\mathrm{ext}\gtrsim 1.7\times 10^6\,\epsilon_\mathrm{e,-1}^{-1}\epsilon_\mathrm{B,-3}^{-1}E_\mathrm{pre,52}^{-1}\Delta t_\mathrm{pre,2.2}\,\mathrm{cm^{-3}}, & k=0;\\
 A_\mathrm{\star}\gtrsim 21\,\epsilon_\mathrm{e,-1}^{-1/2}\epsilon_\mathrm{B,-3}^{-1/2}\Delta t_\mathrm{pre,2.2}\,\mathrm{cm^{-3}}, & k=2.\\
\end{array}
\end{equation}
In the following, we will demonstrate that these conditions are always satisfied in the regions of interest of our parameter space. In the Thomson regime, the IC cooling time scale is
\begin{equation}
t_\mathrm{cool}^\prime\sim\left\lbrace
\begin{array}{lr}
  180\,L_{52.6}^{-1}E_\mathrm{pre,52}^{3/4}n_\mathrm{ext,2}^{-3/4}\Delta t_\mathrm{pre,2.2}^{-1/4}\,\mathrm{s}, & k=0;\\
 14\,L_{52.6}^{-1}E_\mathrm{pre,52}^{3/2}A_\mathrm{\star,0}^{-3/2}\Delta t_\mathrm{pre,2.2}^{1/2}\,\mathrm{s}, & k=2.\\
 
\end{array}\right.
\end{equation}
This is shorter than the illumination phase duration, $t^\prime_\mathrm{cool}\ll \Delta t^\prime_\mathrm{ill}$, with
\begin{equation}
\Delta t_\mathrm{ill}^\prime \gtrsim \Gamma_0 E_\mathrm{\gamma f}/L\approx \left(\frac{E_\mathrm{\gamma f}/L}{40\,\mathrm{s}}\right)\left\lbrace\begin{array}{lr}
  870\,E_\mathrm{pre,52}^{1/8}n_\mathrm{ext,2}^{-1/8}\Delta t_\mathrm{pre,2.2}^{-3/8}\,\mathrm{s} & k=0;\\
 720\,E_\mathrm{pre,52}^{1/4}A_\mathrm{\star,0}^{-1/4}\Delta t_\mathrm{pre,2.2}^{1/4}\,\mathrm{s} & k=2,\\
\end{array}\right.
\end{equation}
provided that
\begin{equation}\label{eq:density_for_IC_cooling}
\begin{array}{lr}
 n_\mathrm{ext}\gg 8\,E_\mathrm{\gamma f,54.2}^{-8/5}E_\mathrm{pre,52}\Delta t_\mathrm{pre,2.2}^{1/5}\,\mathrm{cm^{-3}}, & k=0;\\
 A_\mathrm{\star}\gg 0.04\,E_\mathrm{\gamma f,54.2}^{-4/5}E_\mathrm{pre,52}\Delta t_\mathrm{pre,2.2}^{3/5}\,\mathrm{cm^{-3}}, & k=2,\\
\end{array}
\end{equation}
which is also always verified in our parameter space of interest. We conclude that we can consider the electrons (and any generated pairs) as cold during the illumination phase. This greatly simplifies the computation of the pair enrichment and radiative acceleration.

\subsection{Description of pair enrichment and acceleration}

\begin{figure}
 \centering
 \includegraphics[width=0.75\columnwidth]{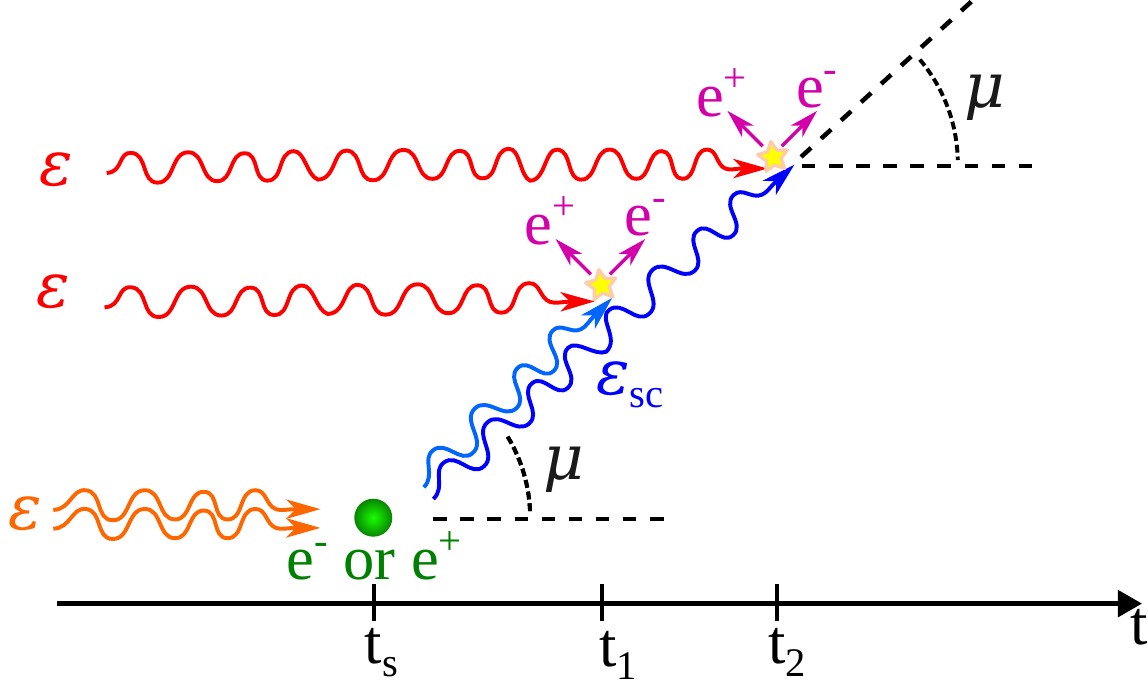}
 \caption{Fundamental block of the process that leads to $e^+e^-$ pair enrichment and acceleration within the blastwave. A packet of incident photons with energy $\varepsilon$ (orange) is inverse-Compton scattered at time $t_\mathrm{s}$ by an electron or positron (green) into a packet of photons with energy $\varepsilon_\mathrm{sc}$ (blue) that propagates with a cosine angle $\cos\theta=\mu$ with respect to the incident photon direction. Subsequent interactions of the scattered photons with incident photons of typical energy $\varepsilon\sim\varepsilon_\mathrm{th}=2/(1-\mu)\varepsilon_\mathrm{sc}$ (red) lead to pair production (two example such interactions happening at times $t_1$ and $t_2$ are shown in the sketch), gradually suppressing the scattered photon packet intensity by a factor $\exp\left[-c(t-t_\mathrm{s})/\lambda_{\gamma\gamma}(\varepsilon_\mathrm{sc},\mu)\right]$, where $\lambda_{\gamma\gamma}$ is the mean free path of the interaction.}
 \label{fig:fundamental_block}
\end{figure}

The basic process that leads to pair production and radiative acceleration in the illumination phase is sketched in Fig. \ref{fig:fundamental_block} and consists of two steps: (i) an incident photon is Compton-scattered off a lepton and (ii) another incident photon annihilates with it, forming an electron-positron pair. Because the blastwave is magnetized, all leptons remain tied to the local plasma. As a result, given that the incident photons are almost perfectly radial, both the scattering and the pair production deposit momentum in the plasma, which is therefore accelerated by the process. The pairs contribute to the optical depth of the medium, enhancing the probability of further scattering of incident photons and consequent pair production. The acceleration, on the other hand, reduces the incident luminosity and redshifts the incident photon spectrum in the comoving frame, working against the former effects. We describe the scattering, pair production and plasma acceleration following a similar treatment as \citet{Beloborodov2002}, hereafter \citetalias{Beloborodov2002}. In Appendix \ref{sec:detailed_model} we give a complete derivation (which allowed us to identify some misprints in the original paper) and technical details about our implementation. Here we mention the fact that we included the feedback of pair annihilation in both the pair production rate and in the acceleration, differently from \citetalias{Beloborodov2002}.

The initial conditions of the calculation are the precursor blastwave parameters, $(A_\star,E_\mathrm{pre})$ or $(n_\mathrm{ext},E_\mathrm{pre})$, the electron fraction $Y_\mathrm{e}$ of the external medium, and the incident spectrum. For the latter, we use the spectral parameters from Section \ref{sec:incident_spec}. The cut-off photon energy $E_\mathrm{cut}$ remains to be specified. Given a set of parameters $(A_\star,E_\mathrm{pre},Y_\mathrm{e},E_\mathrm{cut})$, the calculation yields the pair multiplicity, the pair-enriched blastwave Lorentz factor and the number of annihilated pairs as a function of time, or equivalently of $E_\mathrm{\gamma f}$ (to be understood here as the energy in the incident radiation that flowed through the blastwave until a given time during the illumination phase), $Z_\pm(E_\mathrm{\gamma f})$, $\Gamma_1(E_\mathrm{\gamma f})$ and $N_\mathrm{\pm,ann}(E_\mathrm{\gamma f})$. As discussed in Section \ref{sec:Egf}, we assume that the illumination phase ends at $E_\mathrm{\gamma f}=1.6\times 10^{54}\,\mathrm{erg}$, which corresponds to the main event energy before the steep rise of the LHAASO afterglow.

As we show below, the required pair enrichment can only be reached in presence of a very high external density, which could be linked to enhanced progenitor star mass loss during the late stages of its evolution, as discussed later. The expected electron fraction in such a scenario is $Y_\mathrm{e}=0.5$, and we kept it set to that value (choosing $Y_\mathrm{e}=1$ has a minor impact on our results). For what concerns $E_\mathrm{cut}$, we found its impact to be very limited unless $E_\mathrm{cut}\lesssim 10\,\mathrm{MeV}$, in which case both the enrichment and the radiative acceleration are strongly suppressed. We thus fixed $E_\mathrm{cut}=100\,\mathrm{MeV}$.

\begin{figure}
 \centering
 \includegraphics[width=\columnwidth]{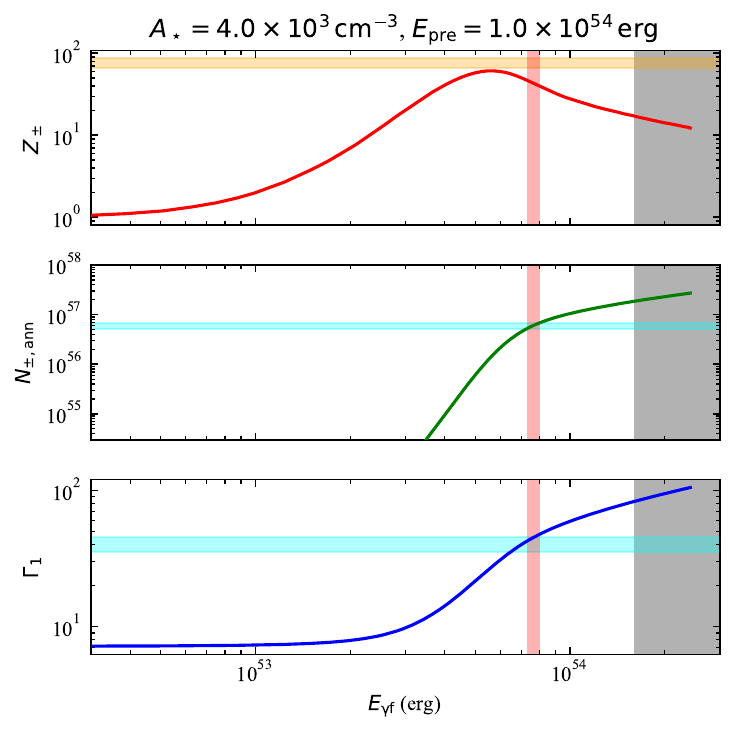}
 \caption{Example illumination phase evolution of pair enrichment and radiative acceleration. The three panels show the evolution of the pair multiplicity ($Z_\pm$, top panel), number of annihilated pairs ($N_\mathrm{\pm,ann}$, middle panel) and bulk Lorentz factor ($\Gamma_1$, bottom panel) as a function of time or equivalently of $E_\mathrm{\gamma f}$ (i.e.\ the total energy in the incident radiation until a given time) for a blastwave in an external medium with $k=2$, with density parameter $A_\star=4\times 10^3\,\mathrm{cm^{-3}}$ and with a precursor blastwave energy $E_\mathrm{pre}=10^{54}\,\mathrm{erg}$, corresponding to an illumination radius $r_0\approx 2.1\times 10^{15}\,\mathrm{cm}$. The grey shaded area corresponds to times after the main event ejecta collided with the blastwave. The vertical red band shows the range of $E_\mathrm{\gamma f}$ for which the number of annihilated pairs satisfy the constraints imposed by the HLE interpretation. The horizontal bands show the pair multiplicity (top panel), number of annihilated pairs (middle panel) and bulk Lorentz factor (bottom panel), respectively, required by the HLE interpretation if the annihilation takes place at $r_0$.}
 \label{fig:example_evolution}
\end{figure}

Figure \ref{fig:example_evolution} shows an example evolution, which corresponds to $k=2$ and $(A_\star,E_\mathrm{pre})=(4\times 10^3\,\mathrm{cm^{-3}},10^{54}\,\mathrm{erg})$. The figure demonstrates that, for these initial parameters, a sufficient number of produced pairs annihilate when the blastwave has the required bulk Lorentz factor.

\subsection{Results}

\begin{figure*}
 \centering
 \includegraphics[width=\columnwidth]{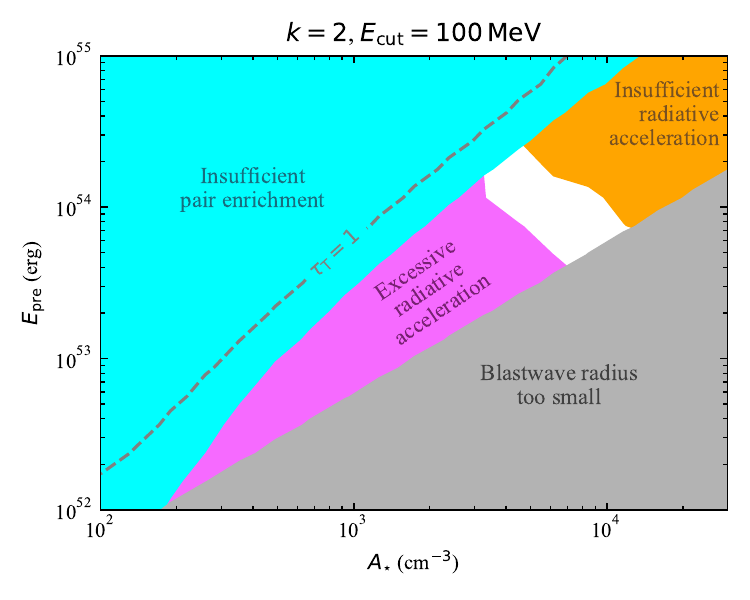}
 \includegraphics[width=\columnwidth]{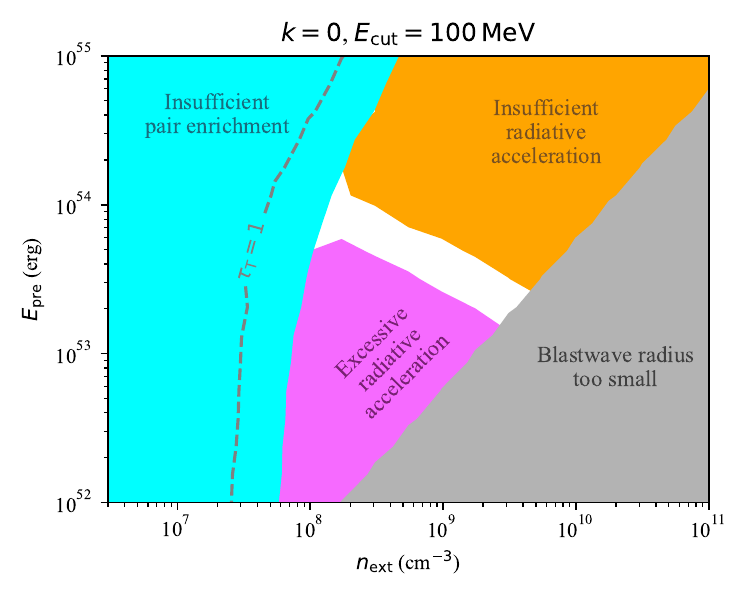}\\
 \caption{Precursor blastwave parameter space constraints assuming $E_\mathrm{cut}=100\,\mathrm{MeV}$. Coloured regions in the figures show regions on the $(A_\star,E_\mathrm{pre})$ plane ($k=2$, left-hand panel) or $(n_\mathrm{ext},E_\mathrm{pre})$ plane ($k=0$, right-hand panel) where the following constraints imposed by the HLE interpretation (see Sect.\ \ref{sec:constraints_from_HLE_fitting}) are violated during the illumination phase: the blastwave radius is smaller than $10^{15}\,\mathrm{cm}$ ($r_{0,16}<0.1$, grey); an insufficient number of annihilating pairs are produced ($N_\mathrm{\pm,ann,57}/r_{0,16}<2.5$, cyan); the radiative acceleration before the pair mass-annihilation is too efficient ($\Gamma_1/r_{0,16}>220$ when $N_\mathrm{\pm,ann,57}/r_{0,16}=2.5$, magenta); the radiative acceleration is insufficient ($\Gamma_1/r_{0,16}<173$ when $N_\mathrm{\pm,ann,57}/r_{0,16}=3.3$, orange). In the region to the right of the dashed line, the blastwave becomes optically thick to Thomson scattering during the illumination phase. }
 \label{fig:param_space_constraints}
\end{figure*}

When varying the initial conditions, we found two possible types of $Z_\pm$ evolution during the illumination phase: either the pair multiplicity peaks and then decreases at some point before the end of the illumination phase (such as in Fig. \ref{fig:example_evolution}), or it increases monotonically. This corresponds to two possible regimes: in the former, the pair density reaches a critical value beyond which the annihilation rate surpasses the pair creation rate, leading to a mass-annihilation during the illumination phase; in the latter, no such mass annihilation takes place, but it can still be triggered by the main event ejecta collision later, as discussed in Appendix \ref{sec:collision_phase}. From our numerical results, we found that all cases which can satisfy the constraints imposed by the HLE interpretation are cases where the mass-annihilation takes place during the illumination phase.

From the analysis described in Section \ref{sec:constraints_from_HLE_fitting}, we found that the number of annihilating pairs should satisfy $2.5<N_\mathrm{\pm,ann,57}/r_{0,16}<3.3$ (union of the 90\% credible intervals of the two analyses discussed there). If this condition was satisfied for some initial conditions, we then checked whether $173<\Gamma_1/r_{0,16}<220$ during the same phase.
Figure \ref{fig:param_space_constraints} shows the constraints imposed by these conditions on the $(A_\star,E_\mathrm{pre})$ or $(n_\mathrm{ext},E_\mathrm{pre})$ plane. The cyan region marks initial conditions for which $N_\mathrm{\pm,ann,57}/r_{0,16}<2.5$ at all times, hence not enough annihilating pairs are produced; the magenta region shows points where enough pairs are produced, but the blastwave Lorentz factor is too high when this happens; similarly, the orange region shows cases for which the blastwave is too slow when the mass-annihilation takes place. The grey region additionally shows cases for which $r_{0,16}<0.1$, that is, the blastwave radius is too small. In the figure we also show a grey dashed line, which marks the initial conditions to the right of which the blastwave \revone{formally} becomes optically thick to Thomson scattering at some point during the illumination phase. This happens consistently when the pair production satisfies our constraints. \revone{We note that this does not constitute per se a problem for the emergence of the annihilation line, because the whole process of pair production and annihilation happens over $\Delta t^\prime_\mathrm{ill}$, which is less than a light-crossing time of the blastwave $\Delta t_\mathrm{lc}^\prime\sim r_0/\Gamma_0 c$. Indeed, it turns out  that the ratio $\Delta t^\prime_\mathrm{ill}/\Delta t^\prime_\mathrm{lc}\sim E_\mathrm{\gamma f}/8 L \Delta t_\mathrm{pre}\sim 3\times 10^{-2}[(E_\mathrm{\gamma f}/L)/40\,\mathrm{s}] \Delta t_\mathrm{pre,2.2}^{-1}$.}

\section{Discussion}
\label{sec:discussion}

For both the $k=0$ and $k=2$ cases, a (quite narrow) region in our parameter space exists where all constraints are satisfied. This corresponds to $10^8\lesssim n_\mathrm{ext}/\mathrm{cm^{-3}}\lesssim 10^{9}$ and $2\times 10^{53}\lesssim E_\mathrm{pre}/\mathrm{erg}\lesssim 10^{54}$ for $k=0$, or $3\times 10^3\lesssim A_\mathrm{\star}/\mathrm{cm^{-3}}\lesssim 10^{4}$ and $4\times 10^{53}\lesssim E_\mathrm{pre}/\mathrm{erg}\lesssim 3\times 10^{54}$ for $k=2$. Qualitatively, these conditions correspond to a very low prompt emission efficiency of the precursor (of about $0.1-1$ percent) and a very high external density, which in both cases is $10^{8}-10^{9}\,\mathrm{cm^{-3}}$ at the distance where the pair-loading takes place. Here we discuss our interpretation of these outcomes.

\subsection{Interpretation of the high required external density} The high external density could be due to a dense circum-stellar medium (CSM) in the close vicinity of the progenitor (our results only require the high density region to extend out to $r_0\sim\mathrm{few}\times 10^{15}\,\mathrm{cm}$). This could be realized if the stellar wind of the progenitor massive star was enhanced during the last few years before the supernova explosion. This scenario is best described assuming $k=2$. Calling $\dot M_\mathrm{w}$ the progenitor mass-loss rate and $v_\mathrm{w}$ the typical velocity of the progenitor winds, we have
\begin{equation}
 A_\star = \frac{\dot M_\mathrm{w}}{4\pi v_\mathrm{w} r_\star^2 m_\mathrm{p}} \approx 10^3 \left(\frac{\dot M_\mathrm{w}}{10^{-2}\mathrm{M_\odot\,yr^{-1}}}\right)\left(\frac{v_\mathrm{w}}{10^{3}\mathrm{km\,s^{-1}}}\right)^{-1}.
\end{equation}
Calling $T\approx 1\,T_{7.5}\,\mathrm{yr}$ the duration of the eruptive phase, the extent of the high-density circum-stellar medium around the progenitor star is $R\sim v_\mathrm{w}T \approx 3\times 10^{15}\,v_\mathrm{w,8}T_{7.5}\,\mathrm{cm}$, which easily accommodates the condition $r_0\lesssim R$.

If the enhanced mass loss phase was interrupted some years before the explosion, pressure waves may have had the time to smooth out the external density gradients, resulting in a more homogeneous medium, perhaps better described by $k=0$. In this case, a density $n_\mathrm{ext}=10^{8}n_\mathrm{ext,8}\,\mathrm{cm^{-3}}$ implies a mass enclosed within $R=3\times 10^{15}\,R_{15.5}\,\mathrm{cm}$ of $M\sim m_\mathrm{p}n_\mathrm{ext}4\pi R^3/3 \approx 10^{-2}\,n_\mathrm{ext,8}R_{15.5}^3\,\mathrm{M_\odot}$. Hence both the $k=0$ and $k=2$ cases require a similar, very intense progenitor mass loss in the final years before the final explosion.

Evidence of circum-stellar media with densities as high as, or even higher than, those found here have been invoked to explain narrow-line core-collapse supernovae (Type IIn SNe, \citealt{Schlegel1990}) such as SN1994W \citep{Chugai2004}. The origin could be eruptive mass ejection in a luminous-blue-variable (LBV) phase of the progenitor taking place in the final stages of the stellar evolution \citep{Kiewe2012}. An intriguing alternative proposed by \citet{Chevalier2012} is a scenario where the progenitor is a binary involving a compact object and a red supergiant that undergo a common envelope phase. By moving through the common envelope, the compact object ejects a large amount of matter, giving birth to the dense CSM. The common envelope phase could end with the compact object merging with the core of the companion, resulting in a short-lived Thorne-Zytkow object \citep{Thorne1977} that eventually explodes.

\begin{figure*}
\centering
 \includegraphics[width=0.9\textwidth]{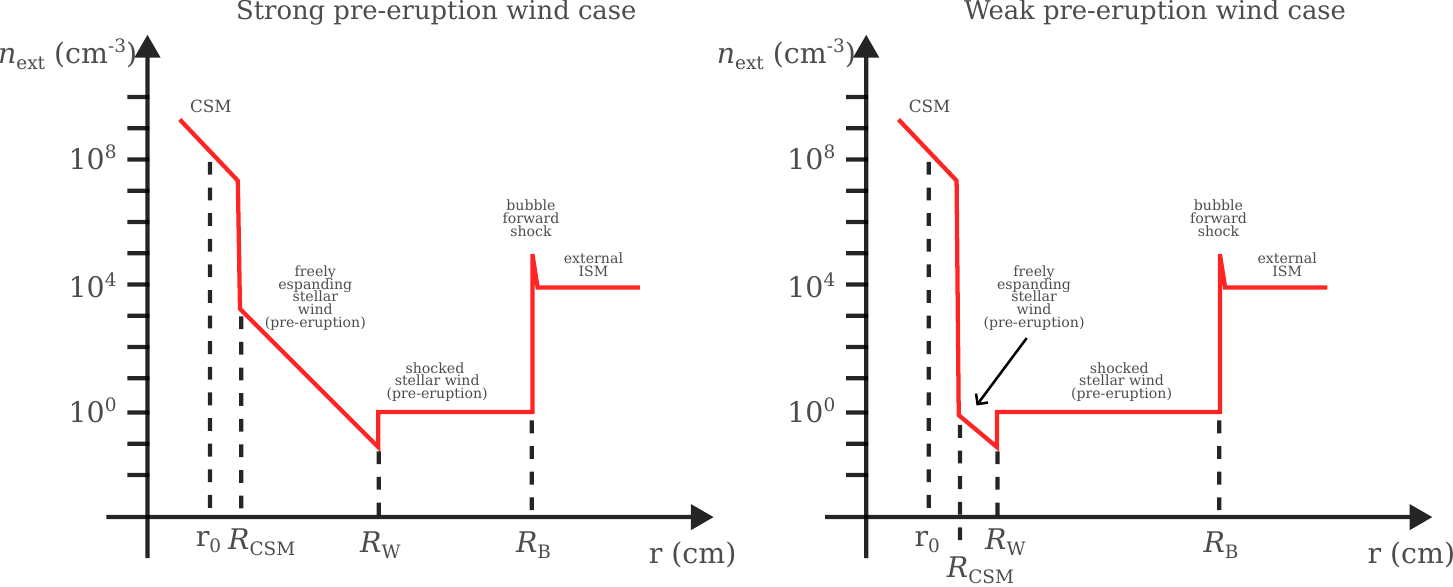}
 \caption{\revtwo{Hypothetical external density profile in two variations of our scenario. The dense CSM occupies a region close to the progenitor, extending out to $R_\mathrm{CSM}\sim \mathrm{few}\times 10^{15}\,\mathrm{cm}$. Outside the CSM, a region occupied by a freely expanding, pre-existing stellar wind could be present. At some radius $R_\mathrm{W}$, which depends on the pre-eruption wind strength and on the external ISM density, such wind reaches its termination shock \citep[][the two panels show a `strong' and a `weak' wind case, with $R_\mathrm{W}$ much closer to $R_\mathrm{CSM}$ in the latter than in the former]{Weaver1977}. The shocked wind has a nearly uniform density profile, which fills the space out to the stellar wind bubble outer radius $R_\mathrm{B}$, where the bubble's forward shock expands into the external ISM.}}
 \label{fig:ext_dens_profile}
\end{figure*}

\revtwo{Figure \ref{fig:ext_dens_profile} shows a sketch of a hypothetical external density profile in our scenario. The dense CSM shell is located close to the progenitor, within a pre-existing stellar wind bubble \citep{Weaver1977,vanMarle2006}. The extent of the bubble and the detailed density profile within the bubble depend primarily on the strength (i.e.\ the kinetic luminosity and the bulk velocity) of the stellar wind and on the density of the external ISM, which likely reflects that of a star forming region \citep{Thakur2026,Chrimes2022}.}

Optical and infrared observations \citep{Fulton2023,Levan2023,Srinivasaragavan2023,Blanchard2024,Kong2024} revealed supernova (SN) emission associated with this GRB, with a luminosity slightly lower than that of the prototypical SN1998bw \citep{Galama1998} and in line with other GRB-associated SNe. This may suggest that, whatever the nature of the progenitor of the peculiar GRB studied here, this should not be vastly different from those of other GRBs with an associated SN, and hence the presence of a CSM could be a common feature of GRB progenitors, rather than a distinctive property of GRB~221009A.

\subsection{Interpretation of the low precursor gamma-ray efficiency}\label{sec:low_precursor_efficiency}
The low gamma-ray efficiency of the precursor contrasts with the very high efficiency of the main event suggested by the main event ejecta Lorentz factor constraint  (Sect.\ \ref{sec:Egf}). A dense CSM, on the other hand, could have acted as a `Thomson screen', letting out only a fraction $e^{-\tau_\mathrm{T,CSM}}$ of the precursor radiation, where $\tau_\mathrm{T,CSM}$ is the CSM optical depth. Under this assumption, we can obtain some instructive constraints on either the outer radius $R_\mathrm{CSM}$ of the CSM (if $k=2$) or the radius $r_\gamma$ of the prompt emission (if $k=0$). In general, the Thomson optical depth of the CSM, for photons emitted at a radius $r_\gamma$, is given by
\begin{equation}
 \tau_\mathrm{T,CSM}\sim \sigma_\mathrm{T}\int_{r_\gamma}^{R_\mathrm{CSM}}n_\mathrm{ext}(r)\,\mathrm{d}r=\frac{\sigma_\mathrm{T}A_\star r_\star}{1-k}\left[\left(\frac{R_\mathrm{CSM}}{r_\star}\right)^{1-k}-\left(\frac{r_\gamma}{r_\star}\right)^{1-k}\right].
\end{equation}
If we require the precursor gamma-ray isotropic-equivalent energy $E_\mathrm{\gamma,pre}\sim 10^{51}\,\mathrm{erg}$ to be equal to $e^{-\tau_\mathrm{T,CSM}}\eta_\mathrm{\gamma,pre}E_\mathrm{pre}$, we have
\begin{equation}
 \tau_\mathrm{T,CSM}\sim \ln\left(\frac{\eta_\mathrm{\gamma,pre}E_\mathrm{pre}}{E_\mathrm{\gamma,pre}}\right)\approx 4.6,
\end{equation}
where the numerical value assumes $E_\mathrm{pre}=10^{54}\,\mathrm{erg}$ and $\eta_\mathrm{\gamma,pre}=0.1$. For $k=2$, assuming $R_\mathrm{CSM}\gg r_\gamma$, we obtain
\begin{equation}
\begin{split}
& r_\gamma \sim \frac{\sigma_\mathrm{T}A_\star r_\star^2}{\ln(\eta_\mathrm{\gamma,pre}E_\mathrm{pre}/E_\mathrm{\gamma,pre})}\approx\\
&\approx 1.7\times 10^{14}A_{\star,3.6}\left[\frac{\ln(\eta_\mathrm{\gamma,pre}E_\mathrm{pre}/E_\mathrm{\gamma,pre})}{4.6}\right]^{-1}\,\mathrm{cm}.
\end{split}
\end{equation}
Conversely, if $k=0$, we obtain a constraint on the CSM radius, which reads
\begin{equation}
\begin{split}
& R_\mathrm{CSM}\sim \frac{\ln(\eta_\mathrm{\gamma,pre}E_\mathrm{pre}/E_\mathrm{\gamma,pre})}{\sigma_\mathrm{T}n_\mathrm{ext}}\approx\\
&\approx7\times 10^{15}n_\mathrm{ext,9}\left[\frac{\ln(\eta_\mathrm{\gamma,pre}E_\mathrm{pre}/E_\mathrm{\gamma,pre})}{4.6}\right]\,\mathrm{cm}.
\end{split}
\end{equation}
Both constraints suggest that the efficiency of the precursor emission could have been `ordinary', but that the observed radiation only consisted of the small fraction of photons that avoided Thomson scattering within the CSM. We give a more detailed account of the effect of a dense CSM on the observed spectrum of the prompt emission in \citet{Oganesyan2026}.

\subsection{Caveat about the Thomson optical depth of the pair-enriched blastwave}
Our calculation of the pair enrichment and acceleration assumes that incident photons undergo only one Compton scattering before annihilating with other incident photons. If the Thomson optical depth of the blastwave becomes larger than unity, multiple scatterings become probable. Intuitively, we expect multiple scatterings to enhance the pair production efficiency beyond what is predicted by our model, because of an enhanced probability of more favourable angles for photon-photon collisions. It is therefore possible that a more advanced calculation could reduce the external density requirements. Nevertheless, because the $\tau_\mathrm{T}=1$ lines in Fig.\ \ref{fig:param_space_constraints} are only slightly to the left of the allowed regions we found with our current treatment, we expect the impact of such a more advanced calculation not to change qualitatively the conclusions of this work.

\subsection{Non-thermal emission by pairs before their annihilation}
Pairs produced during the illumination phase are born mildly relativistic, with $\gamma_\pm\sim 10$ (\citepalias{Beloborodov2002}), before cooling over a time scale $t^\prime_\mathrm{cool}$. The cooling is dominated by IC: the synchrotron emission over the same time scale is comparatively much weaker, as can be seen by taking the ratio of the radiation energy density at the start of the illumination phase $u_\mathrm{rad}^\prime = L/4\pi r_0^2 \Gamma_0^2 c$ to the magnetic energy density in the blastwave, $u_\mathrm{B}^\prime=B^2/8\pi$, which gives, for both $k=0$ and $k=2$,
\begin{eqnarray}
    \frac{u^\prime_\mathrm{rad}}{u^\prime_\mathrm{B}}\approx 0.23 \frac{\Delta t_\mathrm{pre}L}{E_\mathrm{pre}\epsilon_\mathrm{B}}\approx 1.4\times 10^3 \Delta t_\mathrm{pre,2.2}L_{52.6}\epsilon_\mathrm{B,-3}^{-1}E_\mathrm{pre,54}^{-1}.
\end{eqnarray}
The energy released in IC by each pair in the shell comoving frame is simply $2(\gamma_\pm-1)m_\mathrm{e}c^2$. Because the process happens on the same time scale as the mass-annihilation, the luminosity is set by the HLE effect and is therefore approximately equal to the total energy emitted, $E_\mathrm{\gamma,IC}\sim 2(\gamma_\pm-1)\Gamma_\mathrm{HLE}N_\pm m_\mathrm{e}c^2$, divided by the angular time scale $t_\mathrm{ang}\sim \frac{r_0}{2\Gamma_\mathrm{HLE}^2c}$. This yields an IC luminosity of the cooling pairs of about
\begin{equation}
\begin{split}
   & L_\mathrm{\gamma,IC}\approx \left(\frac{\gamma_\pm-1}{10}\right)L_\mathrm{line,50}\left(\frac{t_\mathrm{line}-t_0}{60\,\mathrm{s}}\right)^{-3}\left(\frac{h\nu_\mathrm{line}}{12.6\,\mathrm{MeV}}\right)^{-6}\left(\frac{1+z}{1.15}\right)^{-3}\times\\
   & \times \left\lbrace\begin{array}{lr}
         9\times 10^{50}\,\Delta t_\mathrm{pre,2.2}^{3/4}E_\mathrm{pre,54}^{3/4}n_\mathrm{ext,9}^{-3/4}\,\mathrm{erg\,s^{-1}} & k=0;  \\
         4\times 10^{51}\,\Delta t_\mathrm{pre,2.2}^{3/2}E_\mathrm{pre,54}^{3/2}A_\mathrm{\star,4}^{-3/2}\,\mathrm{erg\,s^{-1}} & k=2.
    \end{array}\right.
\end{split}
\end{equation}
This IC spectrum peaks at around $\gamma_\pm^2\varepsilon_\mathrm{peak} m_\mathrm{e}c^2\approx 260\,\gamma_{\pm,1}^2(\varepsilon_\mathrm{peak}/6)\,\mathrm{MeV}$ and would therefore fall into \textit{Fermi}'s Large Area Telescope (\textit{Fermi}/LAT) band. Intriguingly, emission with a luminosity of a few $10^{51}\,\mathrm{erg\,s^{-1}}$ was seen in \textit{Fermi}/LAT, starting at the same time as the LHAASO afterglow \citep[][see e.g.\ their figure 9]{Axelsson2024}. Whether this emission contains a contribution from the kind of process described here, in addition to the afterglow, is hard to assess, but could be worth a detailed investigation.

\subsection{Similar line emission in other GRBs}

An important question is whether a similar line emission process is to be expected in other GRBs, and in which circumstances. 

The process described here requires a high external density. As demonstrated in this work, even with as favourable a spectrum as that of GRB~221009A, the requirement is $A_\star\gtrsim 10^3\,\mathrm{cm^{-3}}$ (we focus hereafter on $k=2$ for simplicity).

The precursor is also a key ingredient: if a cavity within the high-density environment were not present, the illumination phase would be very brief, because the ejecta would collide immediately with the external medium, accelerating it and hence reducing the incident photon energies in the external medium comoving frame. The condition $r_0\gg r_\gamma$ imposes a stringent requirement on the precursor energy and on the duration of the quiescence that separates it from the main event phase, which reads
\begin{equation}
    E_\mathrm{pre}\Delta t_\mathrm{pre}\gg 10^{54}\,A_{\star,4}r_{\gamma,14}^2\,\mathrm{erg\,s}.
\end{equation}
If we assume a low gamma-ray efficiency for the precursor, $\eta_\mathrm{\gamma,pre}=10^{-3}\eta_\mathrm{\gamma,pre,-3}$ (for the reasons discussed in Section \ref{sec:low_precursor_efficiency}), this can be written as
\begin{equation}
    E_\mathrm{\gamma,pre}\Delta t_\mathrm{pre}\gg 10^{51}\,A_{\star,4}r_{\gamma,14}^2\eta_\mathrm{\gamma,pre,-3}\,\mathrm{erg\,s}.
\end{equation}

 At the beginning of the illumination phase, the blastwave must have decelerated to a low enough Lorentz factor so that a large fraction of the main event spectrum is above the pair production threshold. This can be translated into the requirement $(1+z)E_\mathrm{peak}/\Gamma_0\gtrsim m_\mathrm{e}c^2$, or $\varepsilon_\mathrm{peak}\gtrsim \Gamma_0$. This imposes a stringent requirement on the separation between precursor and main event, namely
 \begin{equation}
     \Delta t_\mathrm{pre}\gtrsim 160\,\varepsilon_\mathrm{peak,0.5}^4E_\mathrm{\gamma,pre,50}\eta_\mathrm{\gamma,pre,-3}A_{\star,4}\,\mathrm{s}.
 \end{equation}
This indicates that the most favourable conditions for finding a similar line in other GRBs are the presence of a long quiescence followed by luminous emission with a very hard spectrum.

\revone{
\subsection{Requirements for the jet to break out of the CSM}
}

\revone{Given the massive and extended CSM found here, a relevant question is whether this constitutes a significant obstacle for the jet breakout, and what requirements this sets on the central engine. The typical time $t_\mathrm{b,\star}$ required for a GRB jet to break out of a Wolf-Rayet progenitor star is on the order of a few tens of seconds \citep{MacFadyen1999,Aloy2000,Matzner2003,Bromberg2011}. On the other hand, even if the jet head were relativistic, it would take at least $t_\mathrm{b,CSM}\gtrsim R_\mathrm{CSM}/c \approx 10^5\,R_\mathrm{CSM,15.5}\,\mathrm{s}\gg t_\mathrm{b,\star}$ for breaking out of an envelope as extended as that discussed here. Therefore, in our scenario the vast majority of the jet propagation time is spent within the CSM, and we can focus solely on that phase.}

\revone{The propagation of the jet head is mainly governed by the dimensionless parameter \citep{Matzner2003,Bromberg2011}}

\begin{equation}
 \tilde L \sim \frac{L_\mathrm{j}}{\Sigma_\mathrm{j}\rho c^3},
\end{equation}
\revone{where $L_\mathrm{j}$ is the kinetic luminosity of the jet, $\Sigma_\mathrm{j}$ is the area of the cross-section of the jet head, and $\rho$ is the ambient medium density in the immediate upstream of the jet head forward shock. If the jet is not collimated by the cocoon, then $\Sigma_\mathrm{j}\sim \pi \theta_\mathrm{j}^2 r^2$, where $\theta_\mathrm{j}$ is the jet half-opening angle, otherwise this is an upper limit to the cross-section. Therefore}
\begin{equation}
\tilde L \gtrsim \frac{L_\mathrm{j,iso}}{2\pi A_\star r_\star^2 m_\mathrm{p} c^3} \approx 1.2\times 10^{5}\,L_\mathrm{j,iso,53}A_{\star,4}^{-1},
\end{equation}
\revone{where $L_\mathrm{j,iso}\sim 2L_\mathrm{j}/\theta_\mathrm{j}^2$ is the isotropic-equivalent jet kinetic luminosity, and we assumed a wind-like density profile. Such a large value of $\tilde L$ suggests that the jet is not collimated by its cocoon \citep[$\tilde L>\theta_\mathrm{j}^{-4/3}$ for any reasonable $\theta_\mathrm{j}$, see][]{Bromberg2011}, and that the head propagates at relativistic speed, with a Lorentz factor}
\begin{equation}
 \Gamma_\mathrm{h}\sim \frac{{\tilde L}^{1/4}}{\sqrt{2}}\approx 13\,L_\mathrm{j,iso,53}^{1/4}A_{\star,4}^{-1/4}.
\end{equation}
\revone{For the jet head propagation to be sustained by the jet until breakout, the engine must remain active over a time}
\begin{equation}
 t_\mathrm{eng}> t_\mathrm{b,CSM}(1-\beta_\mathrm{h})\approx \frac{R_\mathrm{CSM}}{2\Gamma_\mathrm{h}^2\beta_\mathrm{h}c} \approx 300\,R_\mathrm{CSM,15.5}A_{\star,4}^{1/2}L_\mathrm{j,iso,53}^{-1/2}\,\mathrm{s},
\end{equation}
\revone{where $\beta_\mathrm{h}=\sqrt{1-\Gamma_\mathrm{h}^{-2}}$ is the jet head speed in units of the speed of light, otherwise the jet would be choked inside the CSM. This is comparable to the observed duration of GRB~221009A \citep{Lesage2023}, which shows that a successful jet breakout in presence of such an extended envelope (provided that $R_\mathrm{CSM}$ is not much larger than $3\times 10^{15}\,\mathrm{cm}$, which is a typical CSM radius when compared to those inferred from interacting type Ib/c supernovae, see e.g.\ \citealt{Brethauer2022}) does not require fine tuning\footnote{\revone{We note that this breakout time is much shorter than what one would derive from the model by \citet{Hamidani2025}. This is because  their formalism is applicable only when the jet is collimated by the cocoon during the propagation through the CSM, which is not the case here.}}. A minimum engine time for breakout close to the GRB duration is expected to produce a final ejecta structure with a significant amount of energy at relatively large angles outside the jet core \citep{Salafia2020,Salafia2022}. Therefore our scenario, at least qualitatively, lends support to studies that conclude that the ejecta in this GRB could have featured a shallow angular energy structure outside of a narrow core, based on afterglow modelling \citep[e.g.][]{O'Connor2023,Gill2023}.}

~

\revone{\subsection{Impact of the extended CSM on the afterglow}}

\revone{In the analysis presented in this work, we connected the sharp rise in the LHAASO light curve with the moment when the main event ejecta hit the inner boundary of the cavity inflated by the precursor blastwave, but we did not otherwise discuss the impact of the CSM on the multi-wavelength afterglow emission under a broader point of view. Here we give a qualitative account of some of the consequences of our scenario on the subsequent afterglow appearance, and we briefly evaluate its consistency with the available observations.}

\revone{At the moment of the collision, a forward shock (FS) and a reverse shock (RS) form. After crossing the pair-loaded bubble, the FS propagates into the remaining part of the CSM that was not yet shocked by the precursor blastwave, and later expands into the lower-density outer medium that surrounds the CSM. These media are pair-enriched by the prompt emission out to a radius $R_\mathrm{load}\sim 5\times 10^{16}\,E_\mathrm{\gamma f,54}^{1/2}\,\mathrm{cm}$, and pre-accelerated out to $R_\mathrm{acc}\sim 2\times 10^{16}\,E_\mathrm{\gamma f,54}^{1/2}\,\mathrm{cm}$, with a strong dependence of both the pair multiplicity and the bulk Lorentz factor on the radius \citep{Beloborodov2002,Beloborodov2005}. The shock therefore travels through a medium with rapidly varying conditions, in terms of density, pair loading and Lorentz factor. Relativistic leptons in the shock downstream can upscatter the incoming prompt emission photons, and this external inverse Compton (EIC) component itself could even dominate the LHAASO light curve \citep{Beloborodov2014,Vurm2014}. Some observed correlation between the MeV light curve shape and undulations in the LHAASO light curve indeed suggest the importance of EIC up-scattering of the MeV photons in the external shock downstream \citep{Zhang2024_MeV_TeV_relation}. In general, in our scenario the LHAASO afterglow is most likely produced while the FS is still propagating within the CSM.}

\revone{When pair loading is not extreme ($Z_\pm\lesssim Y_\mathrm{e}^{-1} m_\mathrm{p}/m_\mathrm{e}$), the dynamics of the FS is essentially insensitive to the pair-loading and pre-acceleration \citep[][due to a compensation between the reduced ram pressure due to pre-acceleration and the increased pressure due to the relativistic pairs]{Beloborodov2014}. Before the RS crosses the ejecta, and assuming the FS is still within the CSM, the FS downstream Lorentz factor can be estimated by balancing the momentum fluxes at the contact discontinuity\footnote{This is essentially the same calculation as for the jet head propagation in the previous section.} (see \citealt{Beloborodov2014} and Appendix \ref{sec:collision_phase}), which yields $\Gamma_\mathrm{CSM}\sim 11\,L_\mathrm{j,iso,53}^{1/4}A_{\star,4}^{-1/4}$. At this speed, the FS reaches the outer boundary of the CSM in an observer-frame time $t_\mathrm{CSM}\sim(1+z)R_\mathrm{CSM}/\Gamma_\mathrm{CSM}^2 c\approx 950\,[(1+z)/1.15] R_\mathrm{CSM,15.5}L_\mathrm{j,iso,53}^{-1/2}A_\mathrm{\star,4}^{1/2}\,\mathrm{s}$. This time is close to the duration of the prompt emission, which is also approximately equal to the time when the RS crosses the ejecta \citep{Kobayashi2000}. For simplicity, therefore, we can assume that the RS crosses the ejecta at the same time as the FS crosses the CSM.}

\revone{When the FS encounters the density drop at the outer boundary of the CSM \revtwo{(see Fig.\ \ref{fig:ext_dens_profile})}, it accelerates such that $\Gamma\propto r^\alpha$ with $\alpha\sim 1$ (this is the maximum value if the density drop is sufficiently sharp \revtwo{and if the shock is not radiative}, otherwise the acceleration can be slower, see \citealt{Shapiro1980,Nava2013,Derishev2024}). The acceleration saturates at a Lorentz factor
\begin{equation}
\Gamma_\mathrm{\infty}\sim \sqrt{E_\mathrm{j,iso}/M_\mathrm{CSM}c^2}\approx 24 E_\mathrm{j,iso,56}^{1/2}A_\mathrm{\star,4}^{-1/2}R_\mathrm{CSM,15.5}^{-1/2},
\end{equation}
where $E_\mathrm{j,iso}$ is the ejecta isotropic-equivalent kinetic energy. This reflects the intuitive expectation that the initial propagation within the CSM should lead to a FS that is relatively slow (on GRB standards) already at small radii.
The FS expansion continues at approximately constant speed into the lower-density external medium, until the FS sweeps a mass comparable to that of the CSM. Assuming the external medium to be homogeneous with $n_\mathrm{outer}=1\,n_\mathrm{outer,0}\,\mathrm{cm^{-3}}$ \revtwo{(more representative of the `weak pre-eruption wind' case sketched in Fig.\ \ref{fig:ext_dens_profile})}, this happens at a much larger radius $R_\mathrm{dec,outer}\sim (A_\star/n_\mathrm{outer})^{1/3}r_\star^{2/3}R_\mathrm{CSM}^{1/3}\approx 2\times 10^{18}\,A_{\star,4}^{1/3}n_\mathrm{outer,0}^{-1/3}R_\mathrm{CSM,15.5}^{1/3}\,\mathrm{cm}$. In the observer frame, this later deceleration happens at a time $t_\mathrm{dec,outer}\sim (1+z)R_\mathrm{dec,outer}/\Gamma_\infty^2 c\sim 1.6\,A_\mathrm{star,4}^{4/3}n_\mathrm{outer,0}^{-1/3}R_\mathrm{CSM,15.5}^{4/3}E_\mathrm{j,iso,56}^{-1}\,\mathrm{d}$. After this transition, the FS should settle on the self-similar deceleration phase \citep{Blandford1976}.}

\revone{For parameters that explain the line properties in our scenario, the transition times $t_\mathrm{CSM}$ and $t_\mathrm{dec,outer}$ might correspond to the putative `jet break' in the LHAASO light curve at around 900 s \citep{LHAASO2023} and to the achromatic transition identified by \citet{O'Connor2023} at around 1 day post-burst. The Lorentz factor and radius at $t_\mathrm{dec,outer}$ (i.e.\ $\Gamma_\infty$ and $R_\mathrm{dec,outer}$) agree with the lower limits at a similar time (3.46 d) estimated by \citet{Laskar2023} based on the equipartition method of \citet{BarniolDuran2013}.}

\revone{The emission after $t_\mathrm{CSM}$ is a superposition of the emission from the cooling pair-enriched CSM shocked shell (which is likely dominated by synchrotron and is generally expected to be decreasing in luminosity and in typical photon energy) and synchrotron emission from the freshly shocked outer medium (which increases in luminosity up to $t_\mathrm{CSM}$ and should then settle on a `standard' afterglow evolution). Whether the described dynamics and reasonable assumptions on the shock microphysics can explain the detailed light curves and spectra of the GRB221009A afterglow is out of the scope of this work and remains to be investigated. Still, we note that studies of the GRB221009A afterglow so far have clearly demonstrated that its evolution does not fit within a standard scenario, so that non-trivial elements such as an uncommon jet structure \citep[e.g][]{O'Connor2023,Gill2023,Sato2023,Ren2024,Zheng2024} or external density profile \citep[e.g.][]{Kusafuka2025}, evolving microphysical parameters \citep[e.g.][]{Foffano2024} or multiple shocks \citep[e.g.][]{Rhodes2024} have been invoked.
}

~

\section{Summary and conclusions}\label{sec:conclusions}

In this paper, we showed that the narrow emission line at around 10 MeV found in GRB221009A spectra by \citet{Ravasio2024} can be successfully interpreted as high-latitude emission from a thin shell of annihilating $e^-e^+$ pairs, provided that the shell radius, Lorentz factor and pair number satisfy specific constraints. We conceived a scenario that can produce a pair-loaded shell with the required properties: before being loaded with pairs, the shell is the blastwave produced by the ejecta of the GRB221009A precursor when expanding into a dense circum-stellar medium. When the blastwave is located at a radius $r_0\sim\mathrm{few}\times 10^{15}\,\mathrm{cm}$, and it is still expanding with a bulk Lorentz factor of $\Gamma_0\sim \mathrm{few}$, it is illuminated by the bright main event in the GRB prompt emission. Incident photons are scattered within the blastwave and annihilate with other incident photons to produce $e^-e^+$ pairs, which end up being two orders of magnitude more abundant than the original electrons in the blastwave. Momentum deposition by scattering and pair creation accelerates and compresses the blastwave, until the pairs mass-annihilate, producing the line. Because the mass-annihilation happens over a time interval much smaller than $r_0/c$, the observed evolution of the line is dominated by the high-latitude effect. Soon after the mass-annihilation, the main event ejecta catch up and collide with the blastwave and ultimately with the unperturbed external medium, abruptly increasing the afterglow luminosity, which explains the sharp initial rise of the LHAASO light curve. Non-thermal emission by the pairs before their annihilation could have also contributed to the emission seen by \textit{Fermi}/LAT at the same time.

The observed properties of the line can be matched in our scenario provided that the circum-stellar medium around the progenitor had a very high density, corresponding to a mass loss rate of about $10^{-3}-10^{-1}\,\mathrm{M_\odot\,yr^{-1}}$ during the last few years before the explosion, similar to what is inferred from studies of Type IIn supernovae. This provides a clue to the nature of the progenitor, the consequences of which are yet to be fully explored. \revtwo{In particular, the impact of the presence of such a massive CSM on the afterglow emission, and whether the observed afterglow of GRB221009A (and potentially other GRBs) can be successfully interpreted within such a scenario, are yet to be investigated in detail.}

The mechanism described in this work might produce observable emission lines in other GRBs whose progenitor is surrounded by a dense circum-stellar medium, if the jet emission features a long quiescence and provided that the GRB emission after the quiescence is very hard. These requirements can guide the search for similar emission lines in other sources.

\begin{acknowledgements}
O.S.S.\ thanks Sergio Campana for helpful discussions. This work has been funded by the European Union-Next Generation EU, PRIN 2022 RFF M4C21.1 (202298J7KT - PEACE). O.S.S.\ acknowledges funding from INAF through grant 1.05.23.04.04. G.G. acknowledges funding from INAF through grant 1.05.23.06.04.
\end{acknowledgements}

\bibliographystyle{aa}
\footnotesize
\bibliography{references}
\normalsize

\begin{appendix}
\onecolumn

\section{Derivation of the high latitude emission effect}\label{sec:HLE_derivation}

Let us assume the source to be a spherically symmetric, geometrically thin shell of comoving thickness $\Delta r^\prime$ expanding with a bulk Lorentz factor $\Gamma$, and let us call $I(\theta)$ the bolometric intensity (surface brightness) at latitude $\theta$ (angle between the normal to the source surface and the axis parallel to the line of sight). Assuming for simplicity the source to be optically thin (the result does not change in the optically thick case), we can relate the intensity to the comoving bolometric emissivity $j^\prime$ by $I(\theta) = \mathcal{D}^4(\theta) j^\prime \Delta r^\prime$ \citep[e.g.][]{Ghisellini2013}, where
\begin{equation}
 \mathcal{D}(\theta) = \frac{1}{\Gamma(1-\beta\cos\theta)}
\end{equation}
is the Doppler factor, and $\beta=\sqrt{1-\Gamma^{-2}}$. The total fluence (i.e.\ time-integrated flux) received by an observer at a luminosity distance $d_\mathrm{L}$, neglecting cosmological redshift terms, is
\begin{equation}
\mathcal{F} = \frac{r^2}{d_\mathrm{L}^2}\iint \Delta t_\mathrm{obs}(\theta) I(\theta) \sin\theta\,\mathrm{d}\theta\,\mathrm{d}\phi,
\end{equation}
where $\Delta t_\mathrm{obs}(\theta) = (1-\beta\cos\theta)\Delta t_\mathrm{em}=\Delta t_\mathrm{em}/\Gamma \mathcal{D}(\theta)$ is the observed emission duration at latitude $\theta$, and we assumed that $r$ does not change appreciably during the emission (because $\Delta t_\mathrm{em}\ll r/c$). We thus have
\begin{equation}
 E_\mathrm{iso} = 4\pi d_\mathrm{L}^2\mathcal{F} = \iint \frac{\mathcal{D}^3(\theta)}{\Gamma} 4\pi r^2 \Delta r^\prime j^\prime \Delta t_\mathrm{em}\sin\theta\,\mathrm{d}\theta\,\mathrm{d}\phi.
\end{equation}
Since $4\pi r^2 \Delta r^\prime = V^\prime$ is the total source volume in the comoving frame, we have $V^\prime j^\prime \Delta t_\mathrm{em} = L^\prime\Delta t_\mathrm{em}/4\pi = L\Delta t_\mathrm{em}/ 4\pi = E/4\pi$, where $L$ is the source luminosity, $E$ is the total emitted energy, and we exploited the relativistic invariance of the emitted\footnote{The \textit{observed} isotropic-equivalent luminosity, on the other hand, is not an invariant.} luminosity $L^\prime = L$. Defining $\mathrm{d}E/\mathrm{d}\Omega=E/4\pi$ as the energy emitted per unit solid angle by the source, we finally have
\begin{equation}
 E_\mathrm{iso} = \iint \frac{\mathcal{D}^3(\theta)}{\Gamma}\frac{\mathrm{d}E}{\mathrm{d}\Omega}\sin\theta\,\mathrm{d}\theta\,\mathrm{d}\phi,
\end{equation}
which corresponds to the result already derived in \citet{Salafia2015}. The differential isotropic equivalent energy contributed by each `ring' of solid angle $\mathrm{d}\Omega(\theta) = 2\pi \sin\theta\,\mathrm{d}\theta$ is therefore
\begin{equation}
 \frac{\partial E_\mathrm{iso}}{\partial \Omega}(\theta) = \frac{\mathcal{D}^3(\theta)}{\Gamma}\frac{\mathrm{d}E}{\mathrm{d}\Omega}.
\end{equation}
When the HLE effect dominates the observed time evolution of the emission, a single such ring contributes to the observed emission at each time.  The arrival time of photons from the latitude $\theta$ of the ring is
\begin{equation}
 t_\mathrm{obs}(\theta)-t_0 = \frac{r}{c}(1-\cos\theta),
\label{eq:tobs}
 \end{equation}
where  $t_0$ is the arrival time of the first photon. This can be solved for $\cos\theta$ and substituted into the Doppler factor definition, leading to
\begin{equation}
\mathcal{D}(t_\mathrm{obs}) = \frac{\Gamma r}{\Gamma^2 c\left(t_\mathrm{obs}-t_0 + r/(1+\beta)\Gamma^2 c\right)}\equiv \frac{(1+\beta)\Gamma}{1+\left(t_\mathrm{obs}-t_0\right)/t_\mathrm{ang} },
\label{eq:Doppler_tobs}
\end{equation}
where we have defined the `angular time scale' $t_\mathrm{ang}=r/(1+\beta)\Gamma^2 c$. This establishes a one-to-one correspondence between each ring and its associated Doppler factor: the observed time evolution also corresponds to a monotonic decrease in Doppler boosting.

Differentiating Eq.\ \ref{eq:tobs} with respect to $\mathrm{d}\Omega=-2\pi\,\mathrm{d}\cos\theta$, we also obtain
\begin{equation}
 \frac{\partial t_\mathrm{obs}}{\partial \Omega} = \frac{r}{2\pi c},
\end{equation}
which encodes the fact that photons emitted over a differential range of latitudes $(\cos\theta,\cos\theta-\sin\theta\mathrm{d}\theta)$ reach the observer over a range of times $(t_\mathrm{obs}(\theta),t_\mathrm{obs}(\theta) + r\sin\theta\mathrm{d}\theta/c)$.
This leads to
\begin{equation}
 L_\mathrm{iso}(t_\mathrm{obs}) = \frac{\partial \Omega}{\partial t_\mathrm{obs}}\frac{\partial E_\mathrm{iso}}{\partial \Omega} = \frac{2\pi c}{r}\cdot\frac{E}{4\pi}\cdot \frac{(1+\beta)^3 \Gamma^3}{\Gamma\left[1+(t_\mathrm{obs}-t_0)/t_\mathrm{ang}\right]^3}.
\end{equation}
Using $\beta\approx 1$ and rearranging terms, we get to the final expression for the HLE luminosity evolution, namely
\begin{equation}
 L_\mathrm{iso}(t_\mathrm{obs}) \sim \frac{2E/t_\mathrm{ang}}{\left[1+\left(t_\mathrm{obs}-t_0\right)/t_\mathrm{ang}\right]^3},
 \label{eq:L_HLE_appendix}
\end{equation}
which holds until $t_\mathrm{obs}-t_0=2r/c$ (see Eq.\ \ref{eq:tobs}), after which the luminosity goes to zero.
This shows that the HLE effect wipes out the information on the intrinsic luminosity of the emission process (and clearly also any evolution on time scales shorter than the angular time scale), producing a nearly constant luminosity over a time $t_\mathrm{ang}$ followed by a rapid decay $L_\mathrm{iso}\propto (t_\mathrm{obs}-t_0)^{-3}$. The initial luminosity is twice the total emitted energy divided by the angular time scale. The above derivation demonstrates that the results presented by \citet{Peer2024}, which are based on the assumption that the observed luminosity and the comoving one are related by $L_\mathrm{iso}=\mathcal{D}^2 L^\prime$, may need to be revised.

The Doppler factor evolution, Eq.\ \ref{eq:Doppler_tobs}, also sets the spectral evolution, since any frequency evolves as
\begin{equation}
h\nu_\mathrm{obs} = \mathcal{D}(t_\mathrm{obs}) h\nu^\prime = \frac{h\nu_\mathrm{obs,0}}{1+(t_\mathrm{obs}-t_0)/t_\mathrm{ang}},
\end{equation}
where $\nu^\prime$ is the comoving frequency and $\nu_\mathrm{obs,0}=(1+\beta)\Gamma \nu^\prime$. If $F_{\nu}(\nu_\mathrm{obs},t_0)$ is the spectrum at time $t_0$, then at a later time
\begin{equation}
 F_\nu(\nu_\mathrm{obs},t_\mathrm{obs}) = \frac{F_\nu\left(\left(1+\left(t_\mathrm{obs}-t_0\right)/t_\mathrm{ang}\right)\nu_\mathrm{obs},t_0\right)}{\left[1+\left(t_\mathrm{obs}-t_0\right)/t_\mathrm{ang}\right]^{2}},
\end{equation}
where the term at the denominator follows from the requirement that $\nu_\mathrm{obs}F_\nu\propto L_\mathrm{iso}$, where the isotropic-equivalent luminosity evolves as predicted by Eq.\ \ref{eq:L_HLE}.
Assuming a power law spectrum, $F_{\nu}(\nu_\mathrm{obs},t_0)\propto \nu_\mathrm{obs}^{-\beta}$, and focussing on $t_\mathrm{obs}\gg t_\mathrm{ang}$, we recover the original result of \citet{Kumar2000}, namely
\begin{equation}
 F_\nu(\nu_\mathrm{obs},t_\mathrm{obs})\propto \nu_\mathrm{obs}^{-\beta}t_\mathrm{obs}^{-(2+\beta)}.
\end{equation}

\section{HLE model fitting}\label{sec:HLE_fitting}
In this appendix we describe the technical details of our procedure for fitting the HLE model to the line properties inferred from the observations.
In order to account for the inherent degeneracies in the model, we re-parametrized it by defining the parameter vector $\vec x = (x_0,x_1,x_2,x_3)$, with
\begin{equation}
\begin{array}{l}
x_0 = r,\\
x_1 = \Gamma/r,\\
x_2 =  N_\pm/r,\\
x_3 = t_0 \Gamma^2 / r\propto t_0/t_\mathrm{ang}.\\
\end{array}
\label{eq:reparametrisation}
\end{equation}
Each line luminosity and central photon energy error bar from the literature has been inferred using data from a relatively long time bin $[t_a,t_b]$. In order to compare the model with these properties, we defined the model mean luminosity as
\begin{equation}
 \left\langle L_\mathrm{line} \right\rangle_{[t_a,t_b]}(\vec x) = \frac{\int_{t_a}^{t_b}L_\mathrm{line}(t_\mathrm{obs},\vec x)\,\mathrm{d}t_\mathrm{obs}}{t_b-t_a},
\end{equation}
and the model mean line photon energy as
\begin{equation}
 \left\langle h\nu_\mathrm{line} \right\rangle_{[t_a,t_b]}(\vec x) = \frac{\int_{t_a}^{t_b}L_\mathrm{line}(t_\mathrm{obs},\vec x)h\nu_\mathrm{line}(t_\mathrm{obs},\vec x)\,\mathrm{d}t_\mathrm{obs}}{\int_{t_a}^{t_b}L_\mathrm{line}(t_\mathrm{obs},\vec x)\,\mathrm{d}t_\mathrm{obs}}.
\end{equation}
Let the $i$-th error bar be characterized by an inferred line luminosity ${L_{i}}_{-\delta L_{i,\mathrm{low}}}^{+\delta L_{i,\mathrm{high}}}$ and central photon energy ${h\nu_{i}}_{-\delta h\nu_{i,\mathrm{low}}}^{+\delta h\nu_{i,\mathrm{high}}}$, where the sub- and super-scripts indicate one-sigma errors. We defined the additive contribution of this error bar to the log-likelihood as
\begin{equation}
 \ln\mathcal{L}_i(\vec x) = -\frac{1}{2}\left\lbrace\left[\frac{\left\langle L_\mathrm{line} \right\rangle_{[t_a,t_b]}(\vec x)-L_{i}}{\delta L_i(\vec x)}\right]^2 + \left[\frac{\left\langle h\nu_\mathrm{line} \right\rangle_{[t_a,t_b]}(\vec x)-h\nu_{i}}{\delta h\nu_i(\vec x)}\right]^2\right\rbrace,
\end{equation}
where
\begin{equation}
 \delta L_i(\vec x)=\left\lbrace
 \begin{array}{lr}
 \delta L_{i,\mathrm{high}} & \mathrm{if}\,\left\langle L_\mathrm{line} \right\rangle_{[t_a,t_b]}(\vec x)\geq L_{i}\\
 \delta L_{i,\mathrm{low}} & \mathrm{otherwise}
 \end{array}\right.,
\end{equation}
and similarly
\begin{equation}
 \delta h\nu_i(\vec x)=\left\lbrace
 \begin{array}{lr}
 \delta h\nu_{i,\mathrm{high}} & \mathrm{if}\,\left\langle h\nu_\mathrm{line} \right\rangle_{[t_a,t_b]}(\vec x)\geq h\nu_{i}\\
 \delta h\nu_{i,\mathrm{low}} & \mathrm{otherwise}
 \end{array}\right..
\end{equation}
For luminosity upper limits, we simply set $L_i$ equal to the 2-sigma upper limit value, $\delta L_{i,\mathrm{high}}=0$ and $\delta L_{i,\mathrm{low}}=\infty$.
We assumed log-uniform priors on all parameters, that is, $\pi(\vec x) = \prod_{i=0}^3 x_i^{-1}$. Therefore, our posterior probability density took the form
\begin{equation}
p(\vec x\,|\,\vec d) \propto \prod_{i=0}^3 x_i^{-1} \exp\left[\sum_{j=1}^{N}\ln\mathcal{L}_j(\vec x)\right],
\end{equation}
where $N$ is the number of considered error bars. We sampled this posterior probability density using the \texttt{emcee} python package \citep{Foreman-Mackey2013}, then we reconstructed the posterior probability densities of the original parameters $\vec x^\prime = (r,\Gamma,N_\pm,t_0)$ by transforming each sample $\vec x_i$ to the original base, $\vec x^\prime_i = T^{-1}(\vec x_i)$, where $T^{-1}$ is the solution to the system of equations \ref{eq:reparametrisation}.

\section{Hydrodynamical acceleration after ejecta collision}
\label{sec:collision_phase}

\begin{figure*}
 \sidecaption
 \includegraphics[width=10cm]{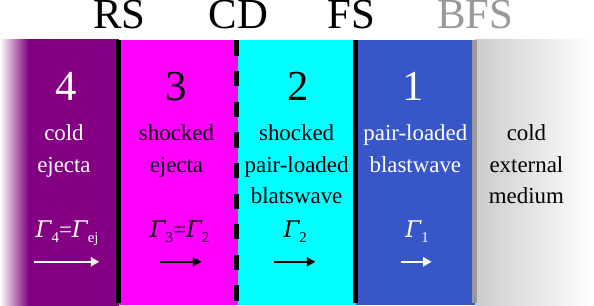}
 \caption{Configuration of the system after the main event ejecta collide with the pair-loaded blastwave, before the forward shock (FS) that arises from the collision crosses the pair-loaded blastwave, catching up with the pre-existing blastwave forward shock (BFS). The reverse shock (RS) separates the cold main event ejecta from the shocked part. A contact discontinuity (CD) separates the ejecta from the the pair-loaded blastwave material.}
 \label{fig:hydro}
\end{figure*}

When the ejecta reach the pair-loaded blastwave, unless the latter has been radiatively accelerated to a bulk Lorentz factor $\Gamma_1\gtrsim \Gamma_\mathrm{ej}$, a violent collision takes place, with the formation of a forward and a reverse shock (RS). The forward shock (FS) first crosses the pair-loaded blastwave, then propagates into the external medium. In the first of these two phases, sketched in  Fig. \ref{fig:hydro}, four regions are relevant to the dynamics: (1) the pair-loaded blastwave, (2) the shocked, pair-loaded blastwave, (3) the shocked ejecta and (4) the un-perturbed ejecta. Regions 1 and 2 are separated by the FS; regions 3 and 4 are separated by the RS; a contact discontinuity (CD) separates regions 2 and 3. Regions 2 and 3 move at the same speed, hence the pressure must balance on the two sides of the contact discontinuity. This can be used \citep[e.g.][]{Sari1995} to derive the bulk Lorentz factor of regions 2-3, as follows.
Let us define the ratio $\ell$ of the gamma-ray luminosity $L$ to the kinetic luminosity $L_\mathrm{j,iso}$ (\revone{both isotropic-equivalent}) of the main event ejecta (after the gamma-rays are released), so that $L = \ell L_\mathrm{j,iso}$. This implies a main event gamma-ray efficiency of $\eta_\gamma = L/(L+L_\mathrm{j,iso})=\ell/(1+\ell)$.  The comoving rest-mass density in region 4 is then $\rho^\prime_4 \sim L_\mathrm{j,iso}/4\pi r_0^2 \Gamma_\mathrm{ej}^2 c^3 = L/4\pi r_0^2 \Gamma_\mathrm{ej}^2 c^3 \ell$.
Calling $\Gamma_2$ the bulk Lorentz factor of regions 2-3, and assuming $\Gamma_2\ll \Gamma_\mathrm{ej}$, the relative Lorentz factor between regions 3 and 4 is $\Gamma_{34}\sim \Gamma_\mathrm{ej}/2\Gamma_2$. The relativistic Rankine-Hugoniot shock jump conditions \citepalias{Blandford1976} then set the pressure in region 3, $p_3\sim 4 \Gamma_{34}^2 \rho^\prime_4 c^2/3$.
Similarly, the pressure in region 2 is $p_2 \sim 4\Gamma_{12}^2 \rho_1^\prime c^2/3$: the relative Lorentz factor can be expressed as $\Gamma_{12}\sim \Gamma_2/2\zeta_\mathrm{racc}\Gamma_0$, where we assumed that prior to the collision the pair loaded shell had a bulk Lorentz factor $\Gamma_1=\zeta_\mathrm{racc}\Gamma_0$. Before being irradiated, the blastwave had a number density (just behind the shock) $n_{1,0}^\prime \sim 4\Gamma_0 n_\mathrm{ext}(r_0)$. After the radiative acceleration, this density becomes  $n_{1}^\prime \sim 4\zeta_\mathrm{racc} \Gamma_0 n_\mathrm{ext}(r_0)$ \citep[][see also Appendix \ref{sec:evol_appendix}]{Madau2000,Beloborodov2002}. Assuming the shell rest mass to be pair-dominated  ($Z_\pm \gtrsim m_\mathrm{p}/2 m_\mathrm{e}$), the density is then $\rho_1^\prime \sim Z_\mathrm{\pm}n_{1}^\prime m_\mathrm{e}$. Requiring $p_2=p_3$, we obtain the Lorentz factor of the shocked pair-loaded shell
\begin{equation}
 \Gamma_2 \sim \left[\frac{L \zeta_\mathrm{racc} \Gamma_0}{16\pi r_0^2 c \ell Z_\pm n_\mathrm{ext}(r_0)m_\mathrm{e} c^2}\right]^{1/4}.
\end{equation}
Imposing $Z_{\pm}=Z_{\pm,\mathrm{HLE}}$ and substituting $r_0$ and $\Gamma_0$ with the expressions in Eqs.\ \ref{eq:r0} and \ref{eq:Gamma0}, we obtain
\begin{equation}
\begin{split}
 &\Gamma_2\approx \left(\frac{(1+z)h\nu_\mathrm{line}}{14.5\,\mathrm{MeV}}\right)^{3/4}L_\mathrm{line,50}^{-1/4} \zeta_\mathrm{racc}^{1/4}L_{52.6}^{1/4}\ell_{-1}^{-1/4}\, \times\\
  &\times \left\lbrace\begin{array}{lr}
  190\, Y_\mathrm{e}^{1/4}E_\mathrm{pre,52}^{1/32}n_{\mathrm{ext},2}^{-1/32}\Delta t_\mathrm{pre,2.2}^{-3/32} & k=0;\\
  155 Y_\mathrm{e}^{1/4} E_\mathrm{pre,52}^{1/16}A_{\star,0}^{-1/16}\Delta t_\mathrm{pre,2.2}^{-1/16} & k=2.\\
  \end{array}\right.
  \end{split}
\end{equation}
If the pair mass-annihilation happens in this phase, then we have also the requirement $\Gamma_2=\Gamma_\mathrm{HLE}$ (Eq.\ \ref{eq:GammaHLE}). In this case, we obtain a constraint on $\ell$, namely
\begin{equation}
\begin{split}
 & \ell \approx \left(\frac{(1+z)}{1.15}\right)^{3}\left(\frac{h\nu_\mathrm{line}}{12.6\,\mathrm{MeV}}\right)^{7}\left(\frac{t_\mathrm{line}-t_0}{60\,\mathrm{s}}\right)^{4}L_\mathrm{line,50}^{-1} L_{52.6}\,\times\\
 & \times \left\lbrace
 \begin{array}{lr}
 3.1\times 10^{-4} \zeta_\mathrm{racc} Y_\mathrm{e} E_\mathrm{pre,52}^{-7/8}n_{\mathrm{ext},2}^{7/8}\Delta t_\mathrm{pre,2.2}^{-11/8}  & k=0;\\
 0.01\, \zeta_\mathrm{racc} Y_\mathrm{e} E_\mathrm{pre,52}^{-7/4}A_{\star,0}^{7/4}\Delta t_\mathrm{pre,2.2}^{-9/4} & k=2.
\end{array}\right.
\end{split}
\end{equation}
When $\ell\ll1$, the above constraint is equivalent to that on the prompt emission efficiency of the main event, $\eta_\gamma$. Large values $\ell>1$ imply large prompt emission efficiencies $\eta_\gamma>0.5$.

These calculations show that, in principle, the mass-annihilation of the pairs (that is, the pair bubble bursting) could happen after the collision between the main event ejecta and the pair-enriched blastwave. If this is the case, the kinetic energy of the main event ejecta must satisfy a constraint in order to accelerate the pair bubble to the Lorentz factor required by HLE.

A relevant question is whether the pairs would annihilate entirely during this phase, or else their annihilation time is longer than the time scale for the shock to cross the pair-loaded shell. A simple calculation shows that the shock crossing time, as measured in the lab frame, is on the order of $r_0/c$. Therefore, the requirement imposed by HLE on the annihilation time scale (that needs be shorter than $r_0/c$) automatically implies that the pairs must annihilate during this phase.

\section{Detailed modelling of the interaction between incident gamma-ray photons and the blastwave}\label{sec:detailed_model}

In this appendix we derive our pair enrichment and radiative acceleration model. The results are very similar to \citetalias{Beloborodov2002}, with some subtle difference, as discussed below.

\subsection{Pair enrichment rate}
\label{sec:pair_enrichment_appendix}

Let us consider a homogeneous shell of radius $r$ expanding at a speed $\beta$ (in units of the speed of light $c$, hence with an associated bulk Lorentz factor $\Gamma$), containing number densities $n_\mathrm{e}$ and $n_\pm$ of cold electrons and pairs, respectively,  that is illuminated by a central source of gamma-rays with a flux $F_{\varepsilon}=L_{\varepsilon}/4\pi r^2$ (here $\varepsilon=h\nu/m_\mathrm{e}c^2$ is an incident photon energy in units of the electron rest mass energy, and $L_{\varepsilon}$ is the specific luminosity of the central source). The specific number density of incident photons is $\mathrm{d}n_\gamma/\mathrm{d}\varepsilon=L_{\varepsilon}/4\pi r^2\varepsilon m_\mathrm{e}c^3=F_{\varepsilon}/\varepsilon m_\mathrm{e}c^3$.

Leptons can Compton-scatter these incident photons. In the lepton rest frame, the differential scattering cross section for incident photons of energy $\varepsilon^\prime = \varepsilon/(1+\beta)\Gamma$ scattered to an angle whose cosine is $\mu^\prime=(\mu-\beta)/(1-\beta\mu)$ is the Klein-Nishina cross section \citep{Jauch1976},
\begin{equation}
\frac{\mathrm{d}\sigma_\mathrm{KN}}{\mathrm{d}\Omega^\prime} = \frac{3\sigma_\mathrm{T}}{16\pi}\cdot\frac{\varepsilon^\prime(1-\mu^\prime)+{\mu^\prime}^2+[1+\varepsilon^\prime(1-\mu^\prime)]^{-1}}{[1+\varepsilon^\prime(1-\mu^\prime)]^2}.
\end{equation}
Since the total cross section is a relativistic invariant, the lab-frame differential cross section is obtained by transforming only the solid angle, so that
\begin{equation}
 \frac{\mathrm{d}\sigma_\mathrm{KN}}{\mathrm{d}\Omega}(\varepsilon,\mu) = \mathcal{D}^2(\mu)\frac{\mathrm{d}\sigma_\mathrm{KN}}{\mathrm{d}\Omega^\prime}(\varepsilon^\prime,\mu^\prime).
\end{equation}
The differential scattering rate density for a single lepton is therefore
\begin{equation}
 \frac{\mathrm{d}^3N_\mathrm{sc}}{\mathrm{d}t \mathrm{d}\varepsilon \mathrm{d}\Omega}(\varepsilon,\mu) =\frac{\mathrm{d}\sigma_\mathrm{KN}}{\mathrm{d}\Omega}(\varepsilon,\mu)(1-\beta)c \frac{\mathrm{d}n_\gamma}{\mathrm{d}\varepsilon}(\varepsilon),
\end{equation}
where the $1-\beta$ term accounts for the relative velocity between the lepton and the incident photons.
In the pre-scattering rest frame of the lepton, the photon energy after scattering, from conservation of momentum and energy, is
\begin{equation}
 \varepsilon^\prime_\mathrm{sc} = \frac{\varepsilon^\prime}{1+(1-\mu^\prime)\varepsilon^\prime}.
\end{equation}
In the lab-frame, this becomes
\begin{equation}
 \varepsilon_\mathrm{sc} = \frac{(1-\beta)\varepsilon}{1-\beta\mu+(1-\mu)\varepsilon/\Gamma}.
\end{equation}
The differential scattering rate density in terms of the scattered photon energy is then
\begin{equation}
 \frac{\mathrm{d}^3N_\mathrm{sc}}{\mathrm{d}t \mathrm{d}\varepsilon_\mathrm{sc} \mathrm{d}\Omega}(\varepsilon_\mathrm{sc},\mu) =\frac{\partial \varepsilon}{\partial \varepsilon_\mathrm{sc}}\cdot\frac{\mathrm{d}\sigma_\mathrm{KN}}{\mathrm{d}\Omega}(\varepsilon,\mu)(1-\beta)c \frac{\mathrm{d}n_\gamma}{\mathrm{d}\varepsilon}(\varepsilon) = \frac{\partial \varepsilon}{\partial \varepsilon_\mathrm{sc}}\cdot\mathcal{D}^2(\mu)\frac{\mathrm{d}\sigma_\mathrm{KN}}{\mathrm{d}\Omega^\prime}\left(\frac{\varepsilon}{(1+\beta)\Gamma},\frac{\mu-\beta}{1-\beta\mu}\right)(1-\beta)\frac{F_{\varepsilon}\left(\varepsilon\right)}{\varepsilon m_\mathrm{e}c^2}.
\end{equation}

Let us now focus on photons scattered within a time $(t_\mathrm{sc},t_\mathrm{sc}+\mathrm{d}t_\mathrm{sc})$ to energies within $(\varepsilon_\mathrm{sc},\varepsilon_\mathrm{sc}+\mathrm{d}\varepsilon_\mathrm{sc})$ and angles within $(\mu,\mu+\mathrm{d}\mu)$, and evaluate their number density at a time $t$. Their initial density is proportional to that of the scattering centers, but reduced by a $(1-\beta)$ factor because the latter move during the time interval $\mathrm{d}t_\mathrm{sc}$, so that
\begin{equation}
 \frac{\mathrm{d}^3 n_\mathrm{sc}}{\mathrm{d}\varepsilon_\mathrm{sc} \mathrm{d}\Omega \mathrm{d}t_\mathrm{sc}}(t=t_\mathrm{sc},t_\mathrm{sc},\varepsilon_\mathrm{sc},\mu)\mathrm{d}\varepsilon_\mathrm{sc}\mathrm{d}\Omega \mathrm{d}t_\mathrm{sc}\sim (1-\beta)\left[n_\mathrm{e}(t_\mathrm{sc})+2 n_\pm(t_\mathrm{sc})\right]\frac{\mathrm{d}^3N_\mathrm{sc}}{\mathrm{d}t\mathrm{d}\varepsilon_\mathrm{sc} \mathrm{d}\Omega}(\varepsilon_\mathrm{sc},\mu,t_\mathrm{sc}) \mathrm{d}\varepsilon_\mathrm{sc} \mathrm{d}\Omega \mathrm{d}t_\mathrm{sc}.
\end{equation}
At a later time $t$, the photons will have travelled to $r(t)=r+\mu c (t-t_\mathrm{sc})$, but since the process of interest happens over a time $t-t_\mathrm{sc}\ll r/c$ (which is also a condition for our planar treatment to remain strictly valid), this dilution can be ignored. Still, an additional process modifies the photon number density with time: during their flight, some of the scattered photons interact with incident photons and annihilate to produce electron positron pairs. Let us indicate with $\lambda_{\gamma\gamma}(\varepsilon_\mathrm{sc},\mu)$ the mean free path of a scattered photon before undergoing such an annihilation. The average pair production time for each of these photons is $\lambda_{\gamma\gamma}/c$, and at time $t$ the number of photons that survived such annihilations is lower by a factor $\exp(-c(t-t_\mathrm{s})/\lambda_{\gamma\gamma})$ with respect to the initial scattered photons. Hence the differential pair production rate density by the annihilation of these photons at time $t$ is
\begin{equation}
\frac{\mathrm{d}^3 \dot n_\mathrm{\pm,\gamma\gamma}}{\mathrm{d}\varepsilon_\mathrm{sc} \mathrm{d}\Omega \mathrm{d}t_\mathrm{sc}}(t,\varepsilon_\mathrm{sc},\mu) \sim \frac{\mathrm{d}^3 n_\mathrm{sc}}{\mathrm{d}\varepsilon_\mathrm{sc} \mathrm{d}\Omega \mathrm{d}t_\mathrm{sc}}(t,t_\mathrm{sc},\varepsilon_\mathrm{sc},\mu)\frac{c}{\lambda_{\gamma\gamma}(\varepsilon,\mu)} = (1-\beta)\left[n_\mathrm{e}(t_\mathrm{sc})+2n_\pm(t_\mathrm{sc})\right]\frac{\mathrm{d}^3N_\mathrm{sc}}{\mathrm{d}t\mathrm {d}\varepsilon_\mathrm{sc} \mathrm{d}\Omega}(\varepsilon_\mathrm{sc},\mu,t_\mathrm{sc}) \mathrm{e}^{-\frac{c(t-t_\mathrm{sc})}{\lambda_{\gamma\gamma}(\varepsilon_\mathrm{sc},\mu)}} \frac{c}{\lambda_{\gamma\gamma}(\varepsilon_\mathrm{sc},\mu)}.
\end{equation}
Setting $n^\star = (n_\mathrm{e}+2n_\pm)(1-\beta)$ and $\mathrm{d}w_\mathrm{sc} = (1-\beta)c\,\mathrm{d}t_\mathrm{sc}=\mathrm{d}(ct_\mathrm{sc}-r)$ and integrating over all scattering times $0<t_\mathrm{sc}<t$ (so that $t=0$ is the time when the shell starts being illuminated), scattering angles and photon energies, after changing one integration variable from $\varepsilon_\mathrm{sc}$ to $\varepsilon$, we obtain the total pair production rate density at time $t$,
\begin{equation}
 \dot n_\mathrm{\pm,\gamma\gamma}(t) = 2\pi \int_{0}^{w(t)}\mathrm{d}w_\mathrm{sc}\int_{-1}^{1}d\mu \,\mathcal{D}^2(\mu,t_\mathrm{sc}) \int_{0}^{\infty}d\varepsilon \frac{F_{\varepsilon}\left(\varepsilon\right)}{\varepsilon m_\mathrm{e}c^2}n^\star(w_\mathrm{sc})\frac{\mathrm{d}\sigma_\mathrm{KN}}{\mathrm{d}\Omega^\prime}\left(\frac{\varepsilon}{(1+\beta)\Gamma},\frac{\mu-\beta}{1-\beta\mu}\right) \frac{\mathrm{e}^{-\frac{w-w_\mathrm{sc}}{\lambda_{\gamma\gamma}(\varepsilon_\mathrm{sc},\mu)}}}{\lambda_{\gamma\gamma}(\varepsilon_\mathrm{sc},\mu)}.
 \label{eq:dotnpm}
\end{equation}
The $\lambda_{\gamma\gamma}$ mean free path is given by \citep{Gould1967}
\begin{equation}
    \lambda^{-1}_{\gamma \gamma}(\varepsilon_\mathrm{sc},\mu) = \int_{\varepsilon_{\rm th}}^{+\infty}(1-\mu) \sigma_{\gamma \gamma}(\varepsilon_\mathrm{sc},\varepsilon,\mu) \frac{\mathrm{d}n_{\gamma}}{\mathrm{d}\varepsilon} \mathrm{d}\varepsilon,
    \label{eq:lmbd_exact_solution}
\end{equation}
where $\varepsilon$ represents again the energy of an incident photon, $\varepsilon_{\rm th}=2/(1-\mu)\varepsilon_\mathrm{sc}$ is the threshold for $\gamma-\gamma$ absorption, and
\begin{equation}
\sigma_{\gamma \gamma}(\varepsilon,\varepsilon_\mathrm{sc},\mu)=\frac{3\sigma_\mathrm{T}}{8x^2}\left[\left(2+2x^{-2}-x^{-4}\right)\ln\left(x+\sqrt{x^{2}-1}\right)-\left(1+x^{-2}\right)\sqrt{1-x^{-2}}\right]
\end{equation}
is the Breit-Wheeler cross section \citep{Jauch1976}, with $x=\sqrt{\varepsilon\varepsilon_\mathrm{sc}(1-\mu)/2}$.

\begin{figure*}
     \sidecaption
 \includegraphics[width=8cm]{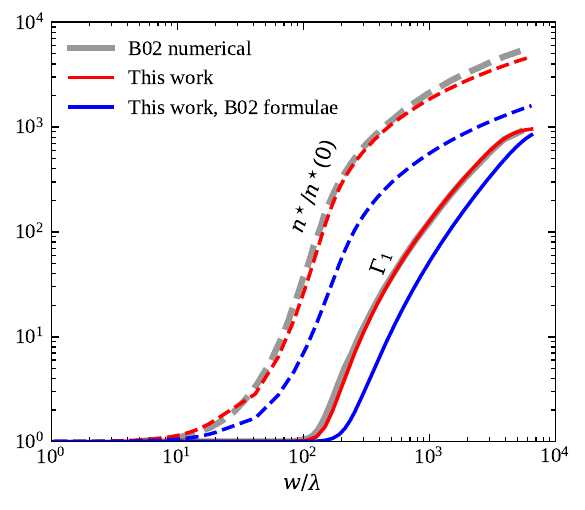}
    \caption{Comparison between our numerical results and those of \citetalias{Beloborodov2002} for a cold and unperturbed shell at a radius $r=10^{14}\,\mathrm{cm}$ illuminated by a central source with $L=10^{51}\,\mathrm{erg/s}$ with a broken power law spectrum peaking at $\varepsilon_\mathrm{p}=1$ and with slopes $\alpha_1=1$ and $\alpha_2=2.5$. Dashed lines show the lepton density evolution ($n^\star/n^\star(0)$), while solid lines show the bulk Lorentz factor $\Gamma_1$ of the illuminated region, as a function of $w/\lambda$, where $\lambda=4\pi r^2 m_\mathrm{e}c^3/\sigma_\mathrm{T}L$ is the electron mean free path for Thomson scattering. Numerical results from \citetalias{Beloborodov2002} (taken from their figure 1) are shown in grey. Results from our model (neglecting pair annihilation) are shown in red. Results obtained using the formulae as written in \citetalias{Beloborodov2002} are shown in blue.\\[20pt]}
    \label{fig:comparison_B02}
\end{figure*}

We note that our Eq.\ \ref{eq:dotnpm} differs from equation 8 of \citetalias{Beloborodov2002} because of an additional $\partial \varepsilon_\mathrm{sc}/\partial \varepsilon$ factor (similarly for Eq.\ \ref{eq:dotPpm}). Also, our Eq.\ \ref{eq:lmbd_exact_solution} contains a $(1-\mu)$ factor that is not present in equation A7 of \citetalias{Beloborodov2002} (but it is present in  \citealt{Gould1967}). Still, our numerical results reproduce very well those of \citetalias{Beloborodov2002} (see the comparison between grey and red lines Fig.\ \ref{fig:comparison_B02}), which suggests that these differences are only due to misprints in that work. This is further supported by the fact that, when using the forms of the equations as given in \citetalias{Beloborodov2002}, we cannot reproduce their results (blue lines in Fig.\ \ref{fig:comparison_B02}).

\subsection{Pair density evolution and radiative acceleration}
\label{sec:evol_appendix}

The auxiliary variable $w=c t- r$ represents the radial distance at time $t$ between the outer edge of the gamma-ray front and the region of interest  \citepalias{Beloborodov2002}. We note that $w(t)= cE_\mathrm{\gamma f}(t)/L$, hence $w L/c$ also represents the amount of radiative energy that crossed the region of interest at time $t$.  In the lab-frame, the number density $n = n_\mathrm{e}+2n_\pm$ evolves with the $w$ variable according to the continuity equation \citepalias{Beloborodov2002}
\begin{equation}
  c\frac{\partial}{\partial w}\left[n(1-\beta)\right] = 2(\dot n_\mathrm{\pm,\gamma\gamma}-\dot n_\mathrm{\pm,ann}),
 \label{eq:continuity_B02}
\end{equation}
where $\dot n_\mathrm{\pm,ann}=3\sigma_\mathrm{T}c n_\pm^2/8\Gamma^2$ is the pair annihilation rate. Differently from \citetalias{Beloborodov2002}, we keep this term as it is not negligible in our context.
Because the source term only affects the pair density, the above equation implies that the density of original electrons satisfies $n_\mathrm{e}^\star(w)=n_\mathrm{e}(w)(1-\beta(w))=n_\mathrm{e}(0)(1-\beta(0))$ and is therefore a conserved quantity \citep{Madau2000}.
In order to describe the radiative acceleration, we exploit the momentum conservation equation \citepalias{Beloborodov2002}
\begin{equation}
 c^2 \frac{\partial}{\partial w}\left[\beta\Gamma\rho(1-\beta) \right] = \dot P_\pm + \dot P_\mathrm{sc} + \dot P_\mathrm{ann},
 \label{eq:mom_conservation}
\end{equation}
where $\rho = Y_\mathrm{e}^{-1}m_\mathrm{p}n_\mathrm{e}+m_\mathrm{e}n$ is the rest-mass density ($Y_\mathrm{e}^{-1}$ being the average number of nucleons per electron in the medium), $\dot P_\pm$ is the momentum deposition rate per unit volume due to pair creation, $\dot P_\mathrm{sc}$ is that due to Compton scattering, and $\dot P_\mathrm{ann}$ is that due to the emission of pair annihilation photons. To compute the first, we assume that the entire momentum of each created pair is deposited into the fluid, due to the presence of a magnetic field in the plasma. The momentum of the pair is equal to that of the annihilating photons, which is $p_\pm\sim (\mu \varepsilon_\mathrm{sc} + \chi\varepsilon_\mathrm{th}(\varepsilon_\mathrm{sc},\mu))m_\mathrm{e}c$, where the quantity $\chi\sim 1$ (which is the ratio of the average absorbed photon energy to the threshold energy) is given in appendix A3 of \citetalias{Beloborodov2002}. Therefore
\begin{equation}
 \dot P_\pm(w) \sim 2\pi\int_{-1}^{1}\int \frac{\mathrm{d}^2 \dot n_{\pm,\gamma\gamma}}{\mathrm{d}\varepsilon_\mathrm{sc}\mathrm{d}\Omega }(\varepsilon_\mathrm{sc},\mu,w) p_\pm(\varepsilon_\mathrm{sc},\mu)\,\mathrm{d}\varepsilon_\mathrm{sc}\mathrm{d}\mu,
\label{eq:dotPpm}
 \end{equation}
where
\begin{equation}
 \frac{\mathrm{d}^2 \dot n_{\pm,\gamma\gamma}}{\mathrm{d}\varepsilon_\mathrm{sc}\mathrm{d}\Omega }(\varepsilon_\mathrm{sc},\mu,w) =\int_0^{w}\frac{\mathrm{d}^3 \dot n_{\pm,\gamma\gamma}}{\mathrm{d}\varepsilon_\mathrm{sc}\mathrm{d}\Omega\mathrm{d}t_\mathrm{s}}(\varepsilon_\mathrm{sc},\mu,t_\mathrm{s}(w_\mathrm{s}),t(w))\frac{\mathrm{d}w_\mathrm{s}}{c}.
\end{equation}

For what concerns the momentum deposition rate due to scattering, we start by computing the difference between the momentum carried by an incident photon, $p_\mathrm{in}=\varepsilon m_\mathrm{e}c$, and the momentum carried by the photon after being scattered, $p_\mathrm{out}= \mu \varepsilon_\mathrm{sc} m_\mathrm{e}c$.
Each scattering hence deposits a momentum $\Delta p = p_\mathrm{in}-p_\mathrm{out}$.
The volumetric momentum deposition rate due to scattering is obtained by integrating this deposited momentum over the volumetric density of scatterings, namely
\begin{equation}
 \dot P_\mathrm{sc}(w)=2\pi \iint \Delta p(\varepsilon_\mathrm{sc},\mu) \frac{\mathrm{d}^3 \dot n_\mathrm{sc}}{\mathrm{d}\varepsilon_\mathrm{sc}\mathrm{d}\Omega\mathrm{d}t_\mathrm{sc}}(w,w_\mathrm{sc}=w,\varepsilon,\mu)\,\mathrm{d}\varepsilon_\mathrm{sc}\mathrm{d}\mu = 2\pi \int_{-1}^{1}d\mu \,\mathcal{D}^2\int_{0}^{\infty}d\varepsilon\,\frac{F_{\varepsilon}\left(\varepsilon\right)}{\varepsilon m_\mathrm{e}c^2}n^\star(w)\frac{\mathrm{d}\sigma_\mathrm{KN}}{\mathrm{d}\Omega^\prime}\left(\frac{\varepsilon}{(1+\beta)\Gamma},\frac{\mu-\beta}{1-\beta\mu}\right) \Delta p(\varepsilon,\mu).
 \label{eq:P_EIC_raw}
\end{equation}

The volumetric momentum loss rate due to the emission of pair annihilation photons, which we include for completeness, is computed as
\begin{equation}
 \dot P_\mathrm{ann}\sim-2\Gamma\beta  m_\mathrm{e}c \dot n_\mathrm{\pm,ann},
\end{equation}
where the approximation adopted stems from the fact that pair annihilation photons are preferentially emitted at $\mu^\prime\sim 0$ in the comoving frame (assuming isotropic pairs), hence $\mu\sim\beta$.

Equations \ref{eq:continuity_B02} and \ref{eq:mom_conservation} can be used to evolve the conservative variables $n_\pm^\star$, $n_\mathrm{e}^\star$ and $\varrho^\star=\beta\Gamma\rho(1-\beta)$. In order to compute in practice the evolution, at each step it is also necessary to compute the primitive variables $n$ and $\Gamma$. To that purpose, we use the relation (which follows from the definitions of $\rho^\star$, $n^\star$)
\begin{equation}
 \beta(w)\Gamma(w) = \frac{\varrho^\star(w)}{Y_\mathrm{e}^{-1}m_\mathrm{p}n_\mathrm{e}^\star(0)+m_\mathrm{e}n^\star(w)},
\end{equation}
from which the bulk Lorentz factor is computed as $\Gamma(w) = \sqrt{1+(\beta(w)\Gamma(w))^2}$.

The initial condition of Eq.\ \ref{eq:continuity_B02}  is (cfr.\ Section \ref{sec:gamma_distributions})
\begin{equation}
 n^\star(0) = n^\star_\mathrm{e}(0) = (4\Gamma_0+3)\Gamma_0 Y_\mathrm{e} n_\mathrm{ext}(r_0)\left(1-\beta_0\right),
\end{equation}
where $\beta_0=\sqrt{1-\Gamma_0^{-2}}$. For Eq.\ \ref{eq:mom_conservation}, the initial condition is
\begin{equation}
 \rho^\star(0)=\beta_0\Gamma_0(Y_\mathrm{e}^{-1}m_\mathrm{p}+m_\mathrm{e})n_\mathrm{e}^\star(0).
\end{equation}

We integrate the differential equations with Euler's method. After the solution is computed, we obtain the pair multiplicity as a function of the gamma-ray energy of the illumination phase,
\begin{equation}
 Z_\pm(E_\mathrm{\gamma f})=\left.\frac{n^\star(w)}{n^\star(0)}\right|_{w=c E_\mathrm{\gamma f}/L},
\end{equation}
and the corresponding bulk Lorentz factor
\begin{equation}
 \Gamma_1(E_\mathrm{\gamma f})=\left.\Gamma(w)\right|_{w=c E_\mathrm{\gamma f}/L}.
\end{equation}
The number of pairs that annihilated is obtained as follows. First, the pair annihilation rate is expressed as
\begin{equation}
 \dot N_\mathrm{\pm,ann} = V\dot n_\mathrm{\pm,ann}=\frac{3\sigma_\mathrm{T}c}{8\Gamma^2}\cdot n_\pm\cdot N_\pm =\frac{3\sigma_\mathrm{T}c}{8\Gamma^2}\cdot \frac{n_\pm^\star}{1-\beta}\cdot \frac{Z_\pm}{2} N_\mathrm{e}=\frac{3\sigma_\mathrm{T}c N_\mathrm{e}}{32}\left(1+\beta\right)\left(n^\star-n_\mathrm{e}^\star\right)Z_\pm.
\end{equation}
The expression is then integrated over $\mathrm{d}t=\mathrm{d}w/(1-\beta)c$ to obtain
\begin{equation}
 N_\mathrm{\pm,ann}(E_\mathrm{\gamma f})=\frac{3\sigma_\mathrm{T} N_\mathrm{e}}{32}\int_0^{cE_\mathrm{\gamma f}/L}\left(1+\beta(w)\right)^2\Gamma^2(w)\left(n^\star(w)-n_\mathrm{e}^\star\right)Z_\pm(w)\,\mathrm{d}w.
\end{equation}

\end{appendix}

\end{document}